\documentclass[11pt,twocolumn]{article}
\usepackage{rotating}
\usepackage{booktabs}
\usepackage{amsmath, amsthm, amsfonts}
\usepackage{bm}
\usepackage{float}
\usepackage{graphicx}
\usepackage{hyperref}
\usepackage[round, numbers, sort&compress]{natbib}
\usepackage[dvipsnames]{xcolor}
\usepackage{caption}
\usepackage{bibentry}
\nobibliography*
\makeatletter
\renewcommand{\@biblabel}[1]{(#1)}
\makeatother

\usepackage{xcolor}  
\usepackage{soul}    
\definecolor{highlight}{RGB}{255, 255, 0}  
\sethlcolor{highlight}  

\usepackage{xcolor}
\usepackage{cleveref}

\makeatletter
\renewcommand{\fnum@figure}{\textbf{Figure~\thefigure}}
\renewcommand{\fnum@table}{\textbf{Table~\thetable}}
\makeatother

\usepackage[total={170mm, 220mm}]{geometry}

\begin{document}

\title{\vspace{-2mm} The Rhythm of Aging:\\
  \Large\textit{Stability and Drift in the Individual Rate of Senescence}
}

\author{
    Silvio C. Patricio\thanks{\texttt{silca@sam.sdu.dk}} \vspace{-.5mm}\\
    \normalsize\textit{Interdisciplinary Center on Population Dynamics}
    \vspace{-1mm}\\
    \normalsize\textit{University of Southern Denmark}
}
\date{}

\twocolumn[

\maketitle
\vspace{-6mm}

\begin{abstract}
Human aging is marked by a steady rise in the risk of dying with age--a process demographers call senescence. Over the past century, life expectancy has risen dramatically, but is this because we are aging slower, or simply starting it later? Vaupel hypothesizes that the pace at which individuals age may be constant, with gains in longevity coming from the delayed onset of senescence rather than its slowing down. We test this idea using a new framework that decomposes the pace of senescence into three components: a biological baseline, a long-term trend, and the cumulative impact of period shocks. Applying this to cohort mortality data above age 80 from 12 countries, we find that once period shocks are accounted for, there is no statistical evidence of a long-term trend, consistent with Vaupel’s hypothesis. Analyses using lower starting ages yield the same qualitative conclusion. Rather than indicating a change in the process that drives senescence, these variations are consistent with echoes of shared historical events. These results suggest that while longevity has shifted, the rhythm of human aging may be conserved.\\
\noindent

\vspace{2mm}


\end{abstract}
\vspace{6mm}

]
\section*{Introduction}

Aging is the gradual decline in physiological functioning--what we see as graying hair, slower steps, and growing vulnerability to illness and injury. Beneath these visible signs lies senescence, the biological process that drives aging. We focus on \emph{actuarial senescence}--the age-related rise in mortality risk--which, in most adult populations, shows an exponential increase in mortality with age, well described by the Gompertz law \cite{gompertz1825nature}. 

In this representation, the Gompertz slope $b$ measures how quickly risk accelerates as deterioration accumulates. Though not a direct biological measure, $b$ is widely used as a proxy for the rate of aging \cite{finch1990longevity}. A higher $b$ means mortality rises more steeply with age; a lower $b$ suggests a slower pace of senescence. Our concern is with the \emph{individual rate of aging}, defined as the rate at which an individual’s mortality risk accelerates with age. For simplicity, we will refer to this as the \emph{rate of aging} or the \emph{pace of senescence}.

Over the last century, more people have survived to older ages: life expectancy is higher; the age at which most deaths occur has moved to older ages; and later life is, for many, healthier \cite{oeppen2002broken, vaupel2021demographic, callaway2025ageing}. We live longer; the question is whether this reflects a \emph{slower} or \emph{later} aging. 

Vaupel framed this as a testable hypothesis: \textit{the rate at which the risk of dying increases with age for humans may be a basic biological constant that is very similar and perhaps invariant across individuals and over time} \cite{vaupel2010}. From this perspective, gains in life expectancy would reflect delayed aging, not a change in the underlying process of senescence. But if the rate of aging is truly changing, it would suggest that the biological processes underlying senescence are more responsive to environmental, behavioral, or historical conditions than previously assumed \cite{finch2004inflammatory, crimmins2011mortality}.

Empirical tests of this hypothesis have yielded mixed findings. One study of Italian cohorts found that estimates of $b$ varied significantly depending on the statistical method used, raising the possibility that apparent changes could reflect model sensitivity rather than shifts in senescence \cite{barbi2003assessing}. Analyses of the aftermath of large mortality shocks--such as famine and wartime captivity--found a flattening of the aging rate at the population level, likely due to selective survival rather than a biological response \cite{zarulli2012mortality, zarulli2013effect}.

Other studies have tested the constancy of $b$ more directly. One analysis rejected the hypothesis that $b$ is constant across countries, sexes, and cohorts, though the observed differences were modest \cite{salinari2014comparing}. A later study suggested that $b$ might even vary with age, rising before leveling off \cite{salinari2020one}. However, their parameter $b$ is a cohort-level quantity shaped by selective disappearance, not the individual rate of aging itself. Besides this conceptual gap, their model does not separate cohort and age effects, which makes it hard to interpret whether the observed variation reflects fluctuations across cohorts or changes in the aging process.

This may suggest that the variations in $b$ could be historically driven. Period events--such as wars, pandemics, and economic crises--strike multiple cohorts at once, just at different ages \cite{vallin2004convergences}, and their lasting consequences can subtly distort the mortality patterns within each exposed cohort through cumulative shifts \cite{horiuchi2003interspecies, zarulli2012mortality}. As a consequence, when we estimate $b$ cohort by cohort, we may be tracing not a pure signal of the aging process, but the lasting effects of these shared historical events. As these shocks accumulate over time, they can produce variations that mimic a change in the slope of mortality, even if the underlying biological rate is constant \cite{zarulli2012mortality, salinari2014comparing, horiuchi1998deceleration}.

These kinds of latent effects accumulate gradually, move in one direction for a while, then turn \cite{nelson1982trends}. These resemble statistical processes, such as random walks, where increments accumulate over time. If period-driven shocks follow this pattern, they could mimic a changing $b$, even when the biology of aging holds steady \cite{yashin2000mortality}. Trajectories of frailty and mortality are also shaped not only by individual biology but by behavior and shared historical conditions \cite{alter1989frailty}.

We ask whether cohort-to-cohort variation in the Gompertz slope reflects a shift in the pace of aging or the effects of period shocks. We decompose $b$ into a baseline and a latent process. We model the latent process as a random walk with drift, the drift being the cohort trend.

A zero drift would suggest that cohort differences in $b$ are consistent with period shocks whose effects persist but cancel on average; a nonzero drift points to a sustained cohort trend. This keeps $b$ interpretable and provides a model-based test of whether gains in longevity reflect \emph{slower} or \emph{later} aging.


\section*{Materials and Methods}

To estimate the pace of senescence, we focus on late-life mortality, where deaths are more likely to reflect intrinsic aging processes than external causes. The analysis is restricted to ages above 80, where non-senescent mortality plays a smaller role and the individual age pattern closely follows the Gompertz form \citep{horiuchi1998deceleration, rau2008continued}.

We model mortality using the gamma-Gompertz framework, which links individual aging dynamics to cohort-level mortality patterns. At the individual level, the hazard of death follows
\begin{equation*}
    \mu(x|Z=z) = z \cdot a \, e^{bx},
\end{equation*}
where $a$ is baseline mortality, $b$ is the Gompertz slope (interpreted as the individual rate of aging), and $z$ captures unobserved frailty. Assuming $Z$ follows a gamma distribution with mean 1 and variance $\gamma$ \citep{vaupel1979impact, steinsaltz2006understanding}, yields to the cohort-level hazard:
\begin{equation}
    \bar{\mu}(x) = \frac{a e^{bx}}{1 + \gamma \frac{a}{b} \left(e^{bx} - 1 \right)} .
\end{equation}

To incorporate period effects, we allow the cohort-specific slope $b_t$ to evolve over time. Specifically, we decompose it into a baseline rate and a latent stochastic process:
\begin{equation*}
    \log b_t = \log b + X_t,
\qquad
X_t = X_{t-1} + \beta + w_t.
\end{equation*}
Here, $X_t$ captures the cumulative impact of shared historical shocks across cohorts. The term $\beta$ represents a persistent drift, indicating whether the rate of aging changes systematically over time. The innovation term $w_t$ represents short-term cohort-to-cohort fluctuations.

We model $w_t \sim \text{Laplace}(0, \sigma_{\text{rw}})$, with $\sigma_{\text{rw}}$ as the scale parameter governing the dispersion of the innovations. The heavier tails of the Laplace distribution allow occasional large shocks--such as wars or pandemics--to be absorbed as isolated deviations rather than interpreted as sustained trends. Thus, $\sigma_{\text{rw}}$ measures the typical magnitude of cohort-to-cohort fluctuations.

Inference is conducted in a Bayesian framework. The key parameter of interest is the drift term $\beta$. If $\beta = 0$, variation in $b_t$ is consistent with accumulated historical shocks but no sustained change in the rate of aging. A nonzero $\beta$ would indicate a persistent directional shift. To assess statistical sensitivity, we also compute a minimum detectable drift threshold, which quantifies how large a sustained trend would need to be to be distinguishable from stochastic variation.

We analyze male and female cohorts from 12 countries using data from the Human Mortality Database \citep{hmd}. Cohorts are included if sufficient age coverage is available above age 80. Detailed descriptions of priors, estimation procedures, diagnostics, and sensitivity analyses are provided in the \emph{SI Appendix}.

\section*{Results}

\begin{figure}[!ht]
    \centering
    \includegraphics[width=\linewidth]{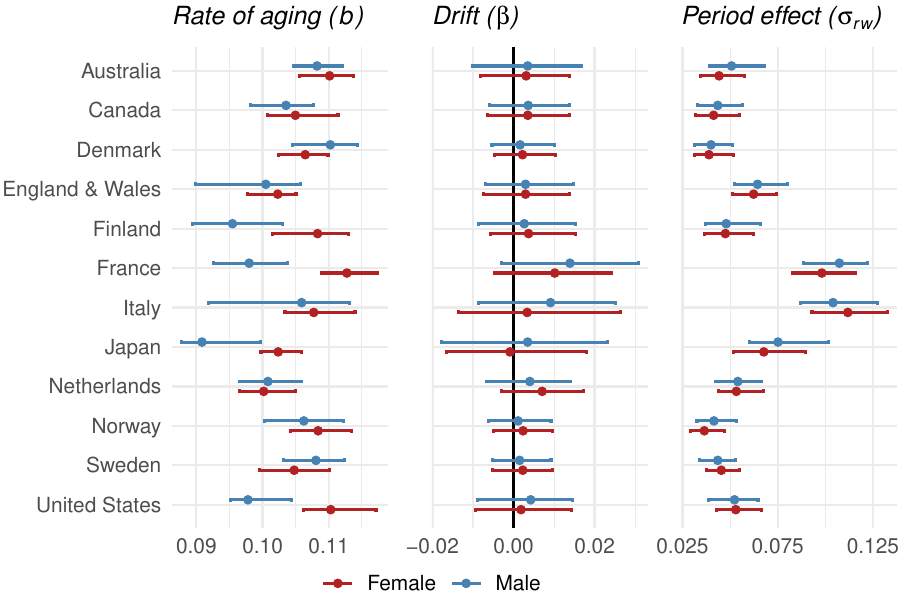}
    \caption{Posterior estimates of the rate of aging ($b$, left), the drift ($\beta$, center), and the variance of the period effect ($\sigma_{\text{rw}}$, right) for males (blue) and females (red) across 12 countries. Points indicate posterior modes, and horizontal bars represent 95\% credible interval. The drift estimates are tightly centered around zero, and the period effect is larger in countries affected by major period shocks, such as France, Italy, and Japan.}
    \label{fig:global_results}
\end{figure}

We applied the random-walk decomposition model to estimate three components of the Gompertz slope--the baseline rate of aging ($b$), the drift term ($\beta$), and the volatility of the period effect ($\sigma_{\text{rw}}$)--for males and females across 12 countries. Figure~\ref{fig:global_results} presents the posterior estimates, and the full numerical results with 95\% highest posterior density intervals are reported in the SI Appendix (Tables S2–S5), together with country-specific decompositions (Figures S2–S5).

Estimates are based on mortality above age 80, where non-senescent mortality plays a minimal role and the age-specific mortality rate is well described by the gamma--Gompertz model \cite{thatcher1998force, horiuchi1998deceleration}. Restricting the analysis to this age range strengthens the interpretation of $b$ as a proxy for the rate of aging, but ties inference to an age threshold.

This restriction reflects the fact that the gamma--Gompertz model is intended to describe senescent mortality. At younger adult ages, mortality reflects a mixture of senescent and non-senescent risks, including causes that do not follow a Gompertz-like trajectory \cite{patricio2024makeham}. As a result, extending the estimation to younger ages introduces a component of mortality that is not captured by the model \cite{steinsaltz2005re}.

However, robustness analyses using lower starting ages (50, 60, and 70; see SI Appendix, Table~S6) lead to the same qualitative conclusion: no sustained drift in the rate of aging. At younger ages, the increased contribution of non-senescent mortality leads to lower estimates of the slope, reflecting a mixture of mortality processes rather than a change in the underlying rate of aging. Details on country coverage under alternative starting ages are provided in the SI Appendix.

\subsubsection*{Cross-Country Consistency in the Rate of Aging}

The estimates of $b$ are relatively consistent across countries. For males, they typically center around 0.102; for females, around 0.107. These differences are small, and in most countries the credible intervals overlap. In the U.S., Japan, and France, however, the male and female intervals do not overlap, suggesting a higher estimated rate of aging for females in those populations.

Countries with larger populations and longer cohort series tend to produce narrower intervals, whereas smaller or shorter series yield wider ones. Precision is also affected by the volatility of period shocks: when cohort-to-cohort fluctuations are large relative to the underlying slope, uncertainty increases, leading to wider credible intervals around the rate of aging.

Across all countries and both sexes, the estimated drift terms remain close to zero, with credible intervals that include zero. Posterior directional indices (Table~S4) likewise indicate weak evidence for sustained change in either direction. Taken together, these results suggest that once stochastic period effects are explicitly modeled, there is no statistical evidence of a persistent directional shift in the Gompertz slope. Differences in precision of both $\beta$ and $b$ primarily reflect variation in cohort length and period volatility, rather than systematic changes in the rate of aging itself.

\subsubsection*{What Drives the Variation?}

The parameter $\sigma_{\text{rw}}$ captures the volatility of the latent period process--that is, the typical magnitude of cohort-to-cohort fluctuations in the estimated $b$ attributable to shocks in calendar time. In practical terms, it summarizes how strongly historical events leave lasting effects on successive cohorts.

Higher $\sigma_{\text{rw}}$ estimates are observed in countries such as France, Italy, and Japan, which experienced major disruptions during the World Wars. Such events can reshape mortality trajectories through demographic mechanisms including selective survival \cite{zarulli2013effect, steinsaltz2006understanding, vaupel1979impact}. Severe shocks disproportionately remove frailer individuals from a cohort, changing the composition of those who survive to older ages.

Because population mortality reflects an average over those who survive, this compositional shift can influence the observed Gompertz slope. The removal of the most vulnerable individuals may leave a more robust surviving group, lowering average mortality at subsequent ages and flattening the observed slope--even if the underlying individual rate of aging has not changed. This interpretation is consistent with the frailty framework, which links mortality deceleration at advanced ages to selection in heterogeneous populations \cite{vaupel1979impact, horiuchi1998deceleration}. In contrast, countries such as the Nordics show lower volatility, consistent with more stable cohort trajectories over time.

\subsubsection*{A Stable Rate Across Countries}

Across the 12 countries, the results follow a consistent pattern. The rate of aging, $b$, remains within a narrow range, whereas the volatility of the period component, $\sigma_{\text{rw}}$, varies in ways that reflect historical and demographic differences. In every country–sex series, the drift parameter $\beta$ is centered close to zero. Credible intervals include zero in all cases, and complementary Bayesian indices reported in the SI Appendix (Table~S4) reinforce this conclusion, providing no evidence that supports a sustained directional trend.  

At the same time, baseline mortality levels ($a_t$) decline across successive cohorts (see Figures S2–S14 in the SI Appendix). Increases in life expectancy therefore could reflect downward shifts in mortality levels rather than systematic changes in the rate at which mortality rises with age.

To evaluate the sensitivity of our framework, we calculated the minimum detectable drift (MDD) for each series (Table~S5, SI Appendix). The detectability thresholds range from 0.76\% to 2.67\% per cohort across countries. In practical terms, the MDD can be interpreted as the smallest sustained percentage change in $b$ from one cohort to the next that would be distinguishable from stochastic variation. The estimated drift parameters fall well below these thresholds, indicating that if a directional trend exists, it must be smaller than what the data can reliably detect.

Sex differences in $b$ persist despite the overall temporal stability. In most countries, females are estimated to have slightly higher values of $b$ than males. One possible mechanism is sex-specific survival selection: males historically experienced higher mortality at younger and middle ages, largely driven by smoking, cardiovascular disease, and external causes such as accidents and violence \cite{beltran2015twentieth}. Stronger early-life mortality selection may produce a more homogeneous group of male survivors at older ages, potentially yielding a flatter observed mortality slope \cite{vaupel1979impact}.

However, these differences are modest and not perfectly consistent across countries. Cross-national variation in smoking histories, war exposure, socioeconomic inequality, and health behaviors may contribute to small differences in estimated $b$ across populations \cite{preston2010new, rogers2010social}. Our model removes stochastic period perturbations but does not account for structural differences in behavioral or epidemiological patterns.

Our analysis does not aim to claim that $b$ is identical across all country–sex populations. Rather, it tests whether $b$ shows a sustained directional change within populations over time. In this respect, for both males and females, the rate of aging shows no statistically significant trend across cohorts. Variation in $b_t$ remains small relative to the volatility introduced by accumulated period shocks, supporting the interpretation that cohort-to-cohort fluctuations primarily reflect historical context rather than changes in the underlying aging process.

\subsection*{Model Diagnostics and Supplementary Estimates}

\begin{figure}[!ht]
    \centering
    \includegraphics[width=\linewidth]{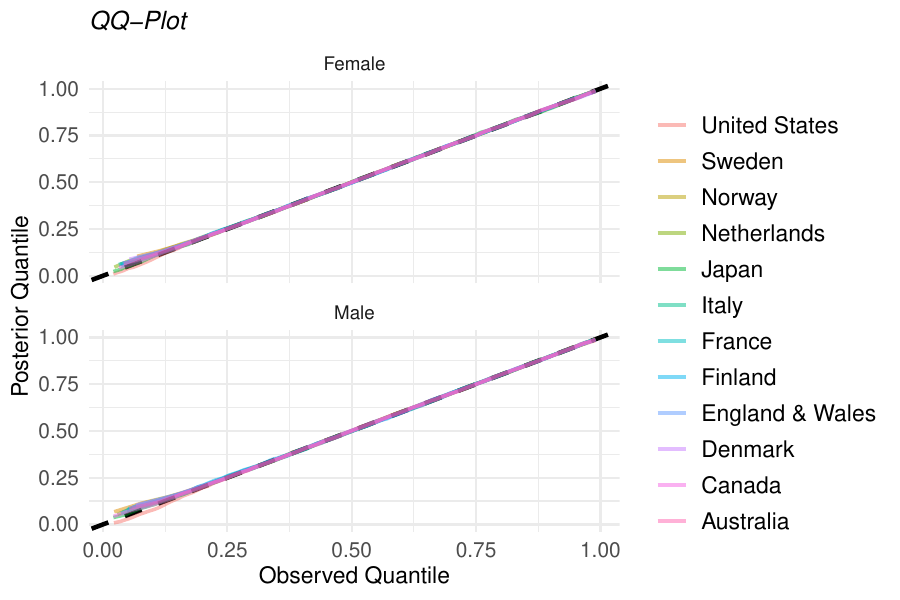}
    \caption{Posterior predictive QQ-plots comparing observed and simulated quantiles of cohort death counts across 12 countries, separately for females (top) and males (bottom). Colored lines show posterior means by country; the black line represents the 45-degree identity line. The close alignment indicates that the model accurately captures the distribution of the observed data across its full range.}
    \label{fig:qqplot}
\end{figure}

Figure~\ref{fig:qqplot} shows the posterior predictive QQ-plots for males and females across all 12 countries. In both cases, the predicted quantiles align closely with the observed ones, falling along the identity line with only minor variation.

Posterior predictive checks evaluate whether the fitted model reproduces the distributional features of the observed death counts \cite{gelman2013bayesian}. Here, the QQ-plots compare the observed data to the distribution implied by the fitted model across the full range of outcomes. The close alignment suggests that the model captures the main structure of the data.

These results support the adequacy of the random-walk specification and the assumption that cohort-level estimates of the Gompertz slope can be explained by baseline, drift, and period components. Country-specific decompositions of these components are provided in \emph{SI Appendix, Supplementary Results and Model Diagnostics}.

\section*{Discussion}

This study revisits Vaupel’s hypothesis that the rate at which the risk of dying increases with age, $b$, is constant. We asked whether cohort fluctuations in $b$ reflect a shift in the pace of aging or the accumulated imprint of shared historical shocks. Once stochastic period variation is accounted for, the estimates of $b$ show no sustained directional trend.

In this framework, $b$ represents the individual pace of senescence as inferred from cohort mortality under the gamma–Gompertz model \cite{vaupel1979impact}. The key question is whether this slope shows sustained directional change across cohorts within a population. After accounting for period shocks, the results are consistent with cohort variation reflecting historical perturbations rather than a systematic change in the aging process.

At the same time, the results do not support the idea of a universally fixed individual rate of aging. Estimated values of $b$ vary modestly across countries and sexes. Although these differences are small, they suggest that the Gompertz slope may not be identical across populations. Because models are estimated separately for each country--sex population, these comparisons should be interpreted descriptively. A formal assessment of cross-population differences would require a joint modeling framework. The evidence therefore favors stability across cohorts within populations rather than a single global constant.

Because our main estimates are based on mortality above age 80, these conclusions speak most directly to the stability of the Gompertz slope in late life, where non-senescent risks play a smaller role. The results therefore pertain to the tempo of aging at advanced ages. They do not exclude the possibility that improvements in earlier-adult survival may shift the timing or level of mortality without immediately changing the individual rate of aging after age 80. Consistent with this interpretation, analyses using lower starting ages (50, 60, and 70) yield the same qualitative conclusion regarding drift (SI Appendix, Table~S6).

This reflects the fact that the interpretation of the Gompertz slope ($b$) as a proxy for the rate of senescence depends on the age range considered and is most direct at ages where mortality is predominantly driven by senescent processes.

\subsubsection*{Support from Past Studies}

Our findings align with Vaupel’s original hypothesis \cite{vaupel2010}, while offering a reinterpretation of previous results. What has been described as model sensitivity \cite{barbi2003assessing} or biological fluctuation \cite{zarulli2012mortality, salinari2014comparing} may instead reflect the cumulative impact of shared historical shocks. Because mortality trajectories are shaped by social and historical environments, cohort-level changes in the estimated Gompertz slope may arise from contextual and compositional effects rather than from shifts in the rate of aging itself \cite{alter1989frailty}.

Comparative research provides further support for a stable rate of senescence. Analyses in primates show that the age-dependent senescent component of the Siler model remains stable within species \cite{colchero2021long}. Similar patterns have been documented in other mammals, where environmental conditions shift overall mortality levels while the underlying rate of senescence remains comparatively stable \cite{berg2025variation, lemaitre2020sex}. 

This interpretation is consistent with recent theoretical work on the structure of mortality dynamics. Models of complex systems show that an exponential increase in mortality can emerge naturally from the failure of many interconnected components, such as those found in biological organisms \cite{nielsen2024gompertz, flietner2024unifying}. In related work, mortality arises from the interaction between the exponential accumulation of health deficits and a nonlinear relationship between deficits and death, while the underlying rate of aging remains fixed \cite{hansen2025we}. 

Across these frameworks, shifts in mortality levels can occur without changing the rate at which the risk of dying increases with age. In this sense, our findings are consistent with the idea that the onset of senescence may move, even if its tempo remains stable. This interpretation may also be compatible with the hallmarks of aging framework, which identifies conserved biological processes such as genomic instability and telomere attrition \cite{lopez2023hallmarks}. The relative stability of the Gompertz slope observed here is compatible with a conserved biological substrate underlying age-related deterioration, despite historical variation in mortality levels.

\subsubsection*{Onset vs. Tempo of Senescence}

A key distinction emerging from our analysis is between the onset and the tempo of senescence. Over the past century, deaths have shifted from younger to older ages, and life expectancy has risen substantially \cite{vaupel2021demographic, oeppen2002broken}. This process of compression primarily reflects declines in baseline and background mortality rather than changes in the Gompertz slope \cite{bongaarts2009trends}.

Our findings suggest that these improvements represent a postponement of the aging process rather than a slowing of its intrinsic pace. Recent work shows that the demographic onset of senescence--the age at which mortality driven by aging overtakes extrinsic risks--has shifted to later ages \cite{patricio2024makeham}. Although this measure is not biological in itself, it identifies when age-related risk becomes the dominant force shaping mortality. Together, these findings suggest that the clock of senescence may start later, even if its rhythm remains conserved after its onset.

From this perspective, gains in life expectancy may reflect delayed aging rather than slower aging. A stable slope combined with shifting mortality levels implies that the timing of deterioration is malleable, while the tempo of aging remains comparatively unchanged.

\subsubsection*{Implications for Longevity Forecasting}

Standard mortality forecasting models, such as Lee–Carter, allow the slope of the log-mortality curve to change over time. Because the time index ($k_t$) interacts with the age profile ($b_x$), the entire age pattern of mortality can rotate, implicitly permitting the rate at which mortality rises with age to drift.

Our results suggest that such flexibility may not be necessary. If the Gompertz slope is stable, long-run improvements in life expectancy are unlikely to reflect sustained changes in the rate at which mortality increases with age. Instead, they are more plausibly attributed to shifts in mortality levels and the postponement of senescence. 

Forecasting models could therefore be simplified by treating the slope as stable over the long term, while allowing short-term fluctuations to be captured by stochastic variation. This approach may yield forecasts that are both more parsimonious and more consistent with observed cohort dynamics.
\subsubsection*{Challenges and Directions from Geroscience}

Our findings do not imply that aging is immutable. Rather, they suggest that the pace at which the individual risk of dying increases with age has been stable. Biomarker-based studies report modifiable "pace of aging" measures \cite{belsky2020quantification}, but it remains unclear whether they correspond directly to the slope of the log-mortality curve or instead capture shifts in damage accumulation that influence the timing of senescence.

Interventions such as senolytics aim to reduce accumulated cellular damage \cite{chaib2022cellular}. If effective, such interventions may lower mortality levels by improving health at given ages. Whether they would also change the Gompertz slope is less clear. Even if biological processes at the individual level were modified, translating such changes into a measurable shift in the population-level rate of aging would likely require sustained and widespread effects across cohorts. At present, population mortality data do not show evidence of a change in the slope. An important empirical task, therefore, is to determine whether future improvements change the overall level of mortality or fundamentally modify the rate at which mortality increases with age.

\subsubsection*{Limitations and Extensions}

Our model assumes constant volatility in the period effect across cohorts. Allowing time-varying volatility--such as through a GARCH-type specification--could test whether the magnitude of historical shocks has changed over time. However, such models require substantially longer time series to support stable estimation \cite{hamilton2020time}. In the present setting, the random-walk specification seems sufficient to absorb variation that might otherwise be misinterpreted as a sustained trend.

A related limitation is that the period component is not age-specific. Historical events may affect cohorts differently depending on the age at exposure. Extending the framework to allow age-by-period interactions would be a natural next step, as major historical events may have different consequences depending on the age at exposure. Such extensions would enrich interpretation. They would, however, face the well-known identification challenges inherent in age–period–cohort analysis \cite{holford1983estimation, fienberg1979identification}.

In addition, our estimates are obtained separately for each country–sex population. A joint hierarchical model could help assess whether the modest differences in $b$ reflect substantive structural variation or sampling uncertainty. By allowing partial pooling, such a framework could improve precision and clarify whether the rate of aging varies systematically between populations.

A more substantive limitation is that our approach relies on specifying a starting age above which mortality is assumed to predominantly reflect senescent dynamics. Although this strategy strengthens interpretability of the Gompertz slope as a proxy for the pace of aging, it ties inference to an age threshold. Developing methods that isolate senescent mortality more directly--without requiring a fixed cutoff--would provide a more general and potentially stronger test of Vaupel’s hypothesis. Approaches that explicitly separate senescent from non-senescent components of mortality could allow the rate of aging to be estimated across a wider age range while preserving its interpretation as a measure of intrinsic aging dynamics. More generally, including ages where mortality is not primarily driven by senescence may introduce a mismatch between the model assumptions and observed mortality patterns, which can affect parameter estimates.

\subsubsection*{Extending to Subgroups}

Our findings may also be relevant for the study of health inequalities. Individuals within the same birth cohort are exposed to the same historical events, but the intensity and consequences of those events differ across socioeconomic groups. Because risks linked to war, economic crises, or infectious disease are unevenly distributed \cite{phelan2010social, elo2009social}, unequal exposure to shared shocks may help explain divergence in survival without necessarily implying differences in the rate of aging.

Recent evidence documents widening survival gaps across socioeconomic groups in Denmark \cite{strozza2025socioeconomic} and substantial life expectancy differences in the United States \cite{bergeron2024inequalities}. Whether such disparities reflect differences in the pace at which mortality rises with age or the cumulative consequences of stratified exposures remains an open empirical question. If the rate of aging were stable across socioeconomic groups, this would be consistent with a shared biological tempo; evidence of systematic slope differences would invite further investigation into both social and biological mechanisms.

\subsubsection*{Clarifying What Changes--and What Does Not}

To our knowledge, this is the first study to decompose the variation in $b$ into a latent period effect. In doing so, we provide a framework that aims to separate the stable component of aging from the noise of period events, using a model that is both parsimonious and grounded in demographic theory.


\section*{Conclusion}

This study revisits Vaupel’s hypothesis that the rate of aging is constant across cohorts. After accounting for shared period shocks, we find little support for a sustained directional trend in the Gompertz slope, $b$. The minimum detectable drift estimates suggest that any long-term change would need to exceed modest thresholds to be distinguishable from stochastic variation. The estimated drift parameters fall below these limits. Together, these findings indicate no evidence of a persistent directional change in the individual rate of aging.

This stability does not imply that aging is fixed in all aspects. Over the past century, survival has shifted toward older ages and life expectancy has increased substantially. These improvements may primarily reflect declines in baseline and background mortality rather than persistent changes in the rate at which mortality rises with age. In demographic terms, the onset of senescence may be postponed even if its tempo remains stable.

Our findings distinguish between changes in level and changes in slope. Over time, mortality levels and the timing of death have shifted, but the rate at which mortality increases with age has remained comparatively stable. Apparent shifts in the pace of aging are more plausibly explained by the accumulated imprint of historical conditions on cohort composition than by changes in the intrinsic rate of senescence. By separating long-term trend from stochastic period variation, we can better understand what has--and has not--changed in the story of human longevity.

\section*{Acknowledgments}

I thank Annette Baudisch, Elisabetta Barbi, James Oeppen, Marie-Pier Bergeron-Boucher, and Trifon Missov for their valuable comments on the manuscript. This research was supported by the AXA Research Fund through the “\textit{AXA Chair in Longevity Research}” and by the European Union (ERC, \textit{Born Once – Die Once}, Grant agreement ID 101043983). The views and opinions expressed are solely those of the author and do not necessarily reflect those of the European Union or the European Research Council Executive Agency. Neither the European Union nor the granting authority can be held responsible for them.

{\footnotesize
\bibliography{bibfile}
}

\newpage

\appendix
\newgeometry{total={170mm, 220mm}}

\renewcommand{\thetable}{S\arabic{table}}
\renewcommand{\thefigure}{S\arabic{figure}}
\renewcommand{\thesubsection}{S\arabic{subsection}}
\setcounter{table}{0}
\setcounter{figure}{0}
\setcounter{subsection}{0}

\twocolumn[
\begin{@twocolumnfalse}
\section*{Supplementary Information}
\vspace{2em}
\end{@twocolumnfalse}
]

\section*{Supplementary Materials and Methods}

To estimate the pace of senescence, we rely on the well-established regularity that adult mortality increases approximately exponentially with age, a pattern captured by the Gompertz law \cite{gompertz1825nature}. This empirical regularity provides a natural bridge between biological aging and population mortality data, allowing the aging process to be studied through age-specific mortality patterns \cite{horiuchi1998deceleration, rau2008continued}.

Vaupel's hypothesis is mathematically grounded in the gamma-Gompertz model, where the exponential increase in the force of mortality with age is modified by unobserved individual frailty \cite{vaupel1979impact}. In this framework, even when populations become more heterogeneous, the rate of individual aging remains constant.

We use the gamma-Gompertz framework to represent individual aging dynamics within a heterogeneous population. Each individual is assumed to follow a Gompertz force of mortality:
\begin{equation*}
    \mu(x|Z = z) = z \cdot a e^{bx},
\end{equation*}
where $a$ is the baseline mortality, $b$ is the Gompertz slope, and $z$ is an unobserved frailty term. The model assumes that $Z$ follows a gamma distribution across individuals, with mean 1 and variance $\gamma$ \citep{vaupel1979impact}, which leads to the cohort-level hazard:
\begin{equation}
    \bar{\mu}(x) = \frac{a e^{bx}}{1 + \gamma \frac{a}{b} \left(e^{bx} - 1 \right)} .\label{eq:hazard}
\end{equation}  

This expression describes how the aggregate force of mortality reflects both the exponential rise in individual risk with age and the selective survival mechanism \cite{vaupel1979impact, steinsaltz2006understanding, vaupel1985heterogeneity}. The parameter $b$, assumed to be invariant across individuals within a cohort, represents the underlying rate at which the individual risk increases with age. In this sense, $b$ captures an individual-level process that is reflected in population mortality, and we interpret it as the individual rate of aging within the cohort. This assumption is the foundation of the model and allows for a direct test of whether the rate of aging is constant \cite{vaupel1979impact, steinsaltz2006understanding, vaupel1985heterogeneity}.

\subsection*{Decomposing Variation in the Individual Rate of Aging}

\begin{figure}[htb]
    \centering
    \includegraphics[width=\linewidth]{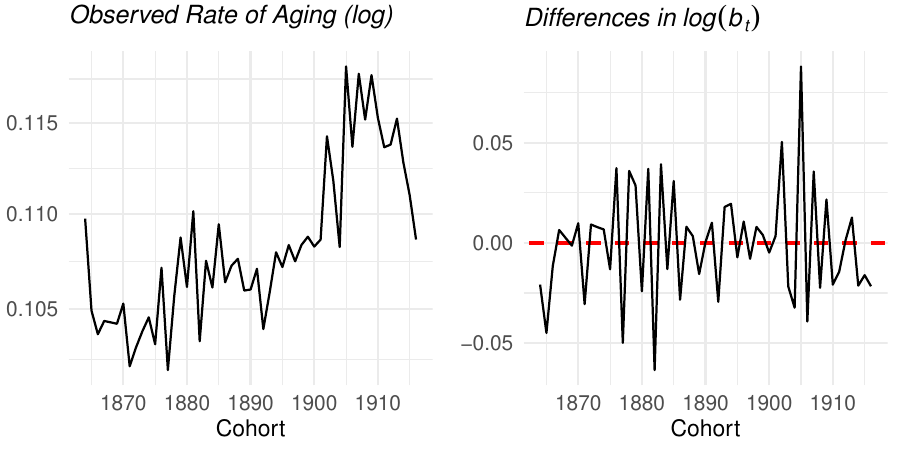}
    \caption{Estimated $b$ across Danish female cohorts and cohort-to-cohort log-difference. The log-difference series is stationary, as confirmed by unit-root and stationarity tests (ADF and PP: $p \leq 0.01$; KPSS: $p \geq 0.10$)}
    \label{fig:motivation}
\end{figure}

When we estimate the rate of aging, $b$, separately for each birth cohort, we often get a series that fluctuates without a clear trend, as illustrated for Danish female cohorts in Figure \ref{fig:motivation} (left panel). While the series itself appears non-stationary, the series of its log-differences (right panel) shows a stationary process \footnote{Stationarity was formally confirmed using a suite of tests at the 5\% significance level. The Augmented Dickey-Fuller (ADF) and Phillips-Perron (PP) tests both strongly reject the null hypothesis of a unit root ($p \leq 0.01$), while the Kwiatkowski-Phillips-Schmidt-Shin (KPSS) test fails to reject the null hypothesis of stationarity ($p \geq 0.10$). Together, these results provide robust evidence that the series of log-differences is stationary.}. This statistical pattern, where a series becomes stationary after differencing, is the classic signature of a random walk.

This statistical structure mirrors the demographic process itself. Historical events like wars, pandemics, or medical breakthroughs can be seen as "shocks" whose effects persist and accumulate, causing the estimated rate of aging to drift unevenly across cohorts. The average of the differences seen in Figure \ref{fig:motivation} (right panel) represents the long-term drift of this process. A drift of zero would suggest that historical shocks cause short-term fluctuations but no lasting directional change in the pace of aging across cohorts. This insight motivates our choice to model these dynamics explicitly.

To formalize this idea, we model the log rate of aging for cohort $t$ as:
\begin{equation}
\log b_t = \log b + X_t,
\end{equation}
with
\begin{equation}
X_t = X_{t-1} + \beta + w_t,
\end{equation}
where:
\begin{itemize}
\item $b$ is the baseline rate of aging,
\item $X_t$ is a latent random walk,
\item $\beta$ is the drift term—capturing any long-run directional change,
\item $w_t$ is white noise with $\mathbb{E}(w_t) = 0$ and variance $\sigma_{\text{rw}}$.
\end{itemize}

\subsubsection*{Parameter interpretation}

In this setup, $X_t$ captures the accumulated impact of historical shocks—a stochastic process that can drift over time. The key implication is that $\mathbb{E}[\log b_t] = \log b + \beta t$: any change in $b_t$ over time comes through the drift. This avoids problems with separating the trend from the cumulative effects and keeps the model simple and interpretable.

Our model is designed to test whether period shocks alone can explain the observed variation in the rate of aging. It is built under the assumption of a fixed biological baseline ($b$), meaning that any persistent directional change across cohorts—whether from cumulative historical effects or a genuine biological trend—is explicitly captured by the drift parameter, $\beta$. It measures whether the cumulative impact of period shocks—both negative (wars, pandemics, economic crises) and positive (medical progress, improved nutrition, sanitation)—pushes the observed rate of aging in a consistent direction over the long run.

The interpretation of $\beta$ is therefore central:
\begin{itemize}
    \item If we find that $\beta = 0$, it implies that period shocks, while causing cohort-to-cohort fluctuations, have no net directional effect over time. This result would provide strong evidence that there is no long-term trend, either historical or biological, in the rate of aging.

    \item If we were to find that $\beta \neq 0$, it would suggest a persistent trend in the rate of aging. For instance, if $\beta < 0$, it would suggest that the net effect of positive historical forces like medical advances consistently outweighs negative shocks, driving the observed rate of aging downward. If $\beta > 0$, it would suggest that the impact of negative shocks has been dominant.
\end{itemize}

This model structure allows us to formally test the parsimonious hypothesis that a single, stable biological rate of aging is sufficient to explain the data once stochastic, non-directional period effects are accounted for. A finding of $\beta \neq 0$ would reject this simple model, indicating a persistent directional force at play. However, it is crucial to note that the model cannot definitively distinguish a purely historical trend from a true change in the underlying biology of senescence. If human biology were evolving toward slower aging, this effect would also be captured in the $\beta$ term. Therefore, while a non-zero $\beta$ would confirm a change in the rate of aging, it could not, by itself, disentangle a historical cause from a biological one.

This structure lets us separate a stable biological rate ($\log b$) from the gradual buildup of shared period effects ($X_t$), and to test whether that accumulation includes a long-term drift ($\beta$). By estimating these components, we can ask whether the changes we see in $b_t$ reflect a true shift in the process that drives aging, or just the long memory of historical shocks.

\subsection*{Model Specification}

To estimate the cohort-specific rate of senescence, we treat observed deaths as outcomes of an underlying stochastic mortality process. Following the classical framework, death counts in large populations can be well approximated by a Poisson process, where deaths are conditionally independent given the force of mortality \cite{brillinger1986biometrics, patricio2020modelagem}. This formulation captures the natural demographic variability of death counts while linking observed events directly to the underlying hazard. Under this framework, we model observed death counts as Poisson-distributed:

\begin{equation*}
D_{x,t} \sim \text{Poisson}(\lambda(x,t) \cdot E_{x,t}),
\end{equation*}
where \( D_{x,t} \) is the number of deaths at age \( x \) for cohort \( t \), \( E_{x,t} \) is the corresponding exposure, and \( \lambda(x, t) \) is the force of mortality, modeled using the gamma-Gompertz hazard in Equation~\ref{eq:hazard}. 

We adopt a Bayesian framework and specify weakly informative priors to regularize estimation while allowing the data to dominate inference:

\begin{eqnarray*}
    a_{t} &\sim& \text{half-Normal}(0, 1) \\
    \gamma_t  &\sim& \text{gamma}(1, 1/2) \\
    X_t|X_{t-1} &\sim& \text{Laplace}(X_{t-1} + \beta, \sigma_{\text{rw}}) \\
    \log b &\sim& \mathcal{N}(0, 2) \\
    \beta &\sim& \mathcal{N}(0, 2) \\
    \sigma_{\text{rw}} &\sim& \text{half-Normal}(0, 1) \\
\end{eqnarray*}

For the positive scale parameters $a_t$ and $\sigma_{\text{rw}}$, we use half-Normal$(0,1)$ priors. These priors enforce positivity while placing most probability mass near smaller values, providing weak but stabilizing regularization \cite{gelman2006prior}.

For the heterogeneity parameter $\gamma_t$, we adopt a Gamma$(1,1/2)$ prior. This specification has been shown to improve identification of frailty variance in gamma–Gompertz models and helps assess whether heterogeneity is supported by the data rather than imposed by the model \cite{patricio2023using}.

The latent cohort process $X_t$ follows a random walk with Laplace innovations. Compared to a Gaussian distribution, the Laplace has heavier tails and a sharper peak at zero. This combination reflects the believe that most cohort-to-cohort changes are small, while reducing sensitivity to occasional large shocks. As a result, rare but substantial historical events are less likely to distort estimation of the drift parameter by being mistaken for persistent directional change. The scale parameter $\sigma_{\text{rw}}$ governs the typical magnitude of these fluctuations and can be interpreted as the volatility of cohort-specific deviations from the baseline aging rate.

For the main parameters of interest, $\log b$ and $\beta$, we use Normal priors centered at zero. These priors impose no prior directional assumption about the baseline rate of aging or its long-term drift, allowing evidence for stability or change to emerge from the data.

\subsection*{Estimation Process}

We implemented the model in a Bayesian framework using \texttt{Stan}, accessed via its R interface \texttt{RStan} \cite{rstan2025}. Four independent Markov chains were run with 6{,}000 iterations each, discarding the first 4{,}000 as warm-up and retaining the remaining 2{,}000 per chain, yielding 8{,}000 posterior draws. Convergence was assessed using rank-normalized $\hat{R}$, which remained below 1.02 for all parameters \cite{vehtari2021rank}. We report posterior modes as point estimates and summarize uncertainty using 95\% highest posterior density (HPD) intervals.

Because the central question concerns whether the drift parameter $\beta$ differs from zero, we complement interval estimates with two posterior summary indices that quantify the compatibility of the data with the null hypothesis of no sustained trend. Table~\ref{tab:estimate_drift} reports the probability of direction ($P$-direction), defined as the posterior probability that $\beta$ is either strictly positive or strictly negative—whichever is more likely \cite{makowski2019indices, van2021cautionary, makowski2019bayestestr}. In practice, it corresponds to the proportion of posterior mass lying on the same side of zero as the posterior median. The index ranges from 0.5 (complete directional uncertainty) to 1 (all mass on one side of zero). For reference, it can be approximately related to a two-sided tail-area measure via $p \approx 2(1 - P\text{-direction})$ \cite{makowski2019indices}.

We also report a maximum a posteriori–based compatibility index ($P$-MAP), defined as the posterior density at $\beta = 0$ divided by the maximum posterior density \cite{makowski2019indices, makowski2019bayestestr}. Unlike $P$-direction, this is not a probability but a relative density measure. Values close to one indicate that the null value lies near the peak of the posterior distribution; values close to zero indicate that it lies in a low-density region. Together, these measures summarize both the directional evidence and the degree to which the null value is supported by the posterior distribution.

\subsection*{Model Validation}

To evaluate model adequacy, we conducted posterior predictive checks based on the alignment between observed and replicated death counts. Rather than relying on a single summary statistic, we assessed model fit using the full quantile structure of the data.

For each observed death count \( y_i \), we computed its empirical quantile and compared it to the posterior predictive distribution of replicated counts \( y_i^{\text{rep}} \). This quantile-wise comparison is used as a diagnostic tool that evaluates whether the posterior distribution reproduces the shape and spread of the observed data. When the model fits well, the posterior quantiles closely track the empirical quantiles, resulting in a QQ-plot that aligns with the identity line. Deviations from this line, such as systematic curvature, would indicate model misfit—for example, underdispersion or structural bias in the predictions.

Such distributional diagnostics are supported in the Bayesian literature as flexible and interpretable checks of model calibration \cite{gelman2000diagnostic, gelman2013bayesian}. They are valuable when the goal is not hypothesis testing but assessing whether the data look plausible under the fitted model—an approach that has been advocated in broader statistical modeling contexts \cite{kruschke2021bayesian}.


\subsubsection*{Sensitivity to Sustained Drift}

To evaluate how sensitive our framework is to a sustained cohort trend, we quantified the minimum detectable drift (MDD) for each country-sex series. Under the random-walk-with-drift specification,
\begin{equation*}
    X_t = X_{t-1} + \beta + w_t,
\end{equation*}
the cohort-to-cohort increments satisfy $\Delta X_t = \beta + w_t$. 

In this setting, the drift $\beta$ is identified through the average increment across cohorts. This approach follows the established logic for detecting deterministic signals within stochastic processes \cite{weatherhead1998factors}. Its statistical precision therefore depends on two elements: the variance of the innovations and the length of the cohort series. As in standard mean estimation, the uncertainty around the average increment decreases at rate $1/\sqrt{n}$, where $n = T-1$ is the number of cohort-to-cohort differences.

In our model, the innovations $w_t$ follow a Laplace distribution with scale $\sigma_{rw}$, implying $\mathrm{Var}(w_t) = 2\sigma_{rw}^2$. Using a large-sample normal approximation for the mean increment, a two-sided $95\%$ detectability threshold is given by
\begin{equation*}
\beta_{\text{MDD}} = 1.96 \cdot \frac{\sqrt{2}\,\sigma_{rw}}{\sqrt{T-1}},
\end{equation*}
where $T$ denotes the number of observed cohorts. This expression reflects a simple principle: a sustained drift must be large relative to the typical stochastic fluctuation, and must accumulate over a sufficiently long series, in order to be distinguishable from noise.

The MDD represents the smallest sustained per-cohort drift that can be separated from stochastic variation, given the volatility and length of the observed series. For interpretability, we translate $\beta_{\text{MDD}}$ into the implied cohort-to-cohort percent change $100\{\exp[\beta_{\text{MDD}}] - 1\}$. Result is present in Table~\ref{tab:MDD}

\subsection*{Data}

We apply our decomposition model to cohort mortality data from the Human Mortality Database \cite{hmd}. The analysis includes male and female birth cohorts from 12 countries: the Nordic countries (Denmark, Finland, Norway, and Sweden), Western and Southern Europe (France, Italy, the Netherlands, and England \& Wales), and selected non-European countries with extensive historical data (Australia, Canada, Japan, and the United States).

To ensure reliable parameter identification, we restricted the analysis to countries with extended and continuous cohort mortality series. Standard time-series diagnostics suggest that non-stationary stochastic processes require a sufficient number of observations—typically at least 50—to distinguish persistent drift from stochastic volatility \cite{box2015time, hamilton2020time}. Because the precision of the drift estimate depends on the signal-to-noise ratio, shorter cohort series provide less leverage to separate sustained change from period-driven fluctuations.

While high-volatility settings (such as France or Italy) may produce wider credible intervals even with long series, shorter cohort series provide less information to separate sustained change from random fluctuation. As a result, estimates for countries with lower cohort coverage (e.g., Japan and the United States) should be interpreted with caution.

To ensure stable estimation of senescence mortality pattern, we include only cohorts with sufficient age coverage. Specifically, we require at least 20 observed age groups above age 80. For robustness, we also estimated models requiring at least 30 age groups above age 70, 40 above age 60, and 50 above age 50. Results for each age threshold are reported in Table~\ref{tab:estimate_different_starting_age}. The substantive conclusions remain unchanged across these alternative truncations, consistently supporting a stable rate of aging. However, data availability at younger ages substantially reduces the number of countries with sufficiently long cohort series, limiting cross-national comparisons under these lower thresholds.

Results presented in the main text focus on mortality above age 80, as this threshold maximizes the number of countries included while concentrating on ages where mortality closely follows the gamma–Gompertz model. The specific cohorts included—by country, sex, and range—are listed in Table~\ref{tab:cohorts}.

\subsubsection*{Materials Sharing}
All code and supporting materials required to reproduce the analysis have been deposited in a public repository at: \url{https://github.com/scpatricio/The_Rhythm_of_Aging_PNAS2026}

\onecolumn

\clearpage   

\section*{Supplementary tables} 
\label{appendix:tables}
This section reports the cohorts included in the analysis and the posterior estimates of the main model parameters, shown with 95\% credible intervals by country and sex.

\begin{table}[H] 
    \centering
    \caption{Cohorts considered in the analysis. Countries, sexes, and ranges of cohorts included.}
    \vspace{-6mm}
    \label{tab:cohorts}
    \begin{tabular}{l l c c c} 
        \\
        \hline
        Country & Sex & First cohort & Last cohort & Length \\
        \hline
        Australia & Male   & 1845 & 1931 & 86\\
                  & Female & 1845 & 1931 & 86\\
        Canada    & Male   & 1840 & 1932 & 92\\
                  & Female & 1840 & 1932 & 92\\
        Denmark   & Male   & 1764 & 1934 & 170\\
                   & Female & 1764 & 1934 & 170\\
               England \& Wales & Male   & 1764 & 1932 & 168\\
      & Female & 1764 & 1932 & 168\\

        Finland   & Male   & 1807 & 1933 & 126\\
                  & Female & 1807 & 1933 & 126\\
        France    & Male   & 1737 & 1932 & 195\\
                  & Female & 1737 & 1932 & 195\\
        Italy     & Male   & 1794 & 1932 & 138\\
                  & Female & 1794 & 1932 & 138\\
        Japan     & Male   & 1868 & 1933 & 65\\
                  & Female & 1868 & 1933 & 65\\
        Netherlands & Male   & 1777 & 1932 & 155\\
                  & Female & 1777 & 1932 & 155\\
        Norway    & Male   & 1770 & 1933 & 163\\
                  & Female & 1770 & 1933 & 163\\
        Sweden    & Male   & 1676 & 1933 & 257\\
                  & Female & 1676 & 1933 & 257\\
        United States & Male   & 1852 & 1933 & 81\\
                  & Female & 1852 & 1933 & 81\\
        \hline
    \end{tabular}
\end{table}

\begin{table}[H]
    \caption{Estimate and 95\% credible interval for the individual rate of aging ($b$)}
        \vspace{-3mm}
    \label{tab:estimate_b}
    \centering
    \begin{tabular}[t]{llccc}
        \hline
        Country & Sex & Estimate & Lower C.I. & Upper C.I.\\
        \hline
        Australia & Male & 0.1080 & 0.1005 & 0.1133\\
& Female & 0.1095 & 0.1062 & 0.1135\\
        Canada & Male & 0.1026 & 0.0972 & 0.1073\\
& Female & 0.1059 & 0.1006 & 0.1099\\
        Denmark & Male & 0.1095 & 0.1031 & 0.1142\\
& Female & 0.1064 & 0.1037 & 0.1087\\
        England \& Wales & Male & 0.0969 & 0.0929 & 0.1034\\
& Female & 0.1006 & 0.0985 & 0.1040\\
        Finland & Male & 0.0963 & 0.0902 & 0.1022\\
& Female & 0.1072 & 0.1022 & 0.1138\\
        France & Male & 0.0999 & 0.0960 & 0.1039\\
& Female & 0.1107 & 0.1066 & 0.1149\\
        Italy & Male & 0.1000 & 0.0961 & 0.1041\\
& Female & 0.1105 & 0.1078 & 0.1132\\
        Japan & Male & 0.0931 & 0.0857 & 0.0979\\
& Female & 0.1043 & 0.0984 & 0.1069\\
        Netherlands & Male & 0.1006 & 0.0971 & 0.1047\\
& Female & 0.1010 & 0.0970 & 0.1053\\
        Norway & Male & 0.1064 & 0.1015 & 0.1115\\
& Female & 0.1091 & 0.1050 & 0.1121\\
        Sweden & Male & 0.1078 & 0.1047 & 0.1114\\
& Female & 0.1047 & 0.1017 & 0.1078\\
        United States & Male & 0.0955 & 0.0930 & 0.1016\\
& Female & 0.1110 & 0.1089 & 0.1130\\
        \hline
    \end{tabular}
\end{table}

\begin{table}[H]
    \caption{Estimate and 95\% credible interval for drift ($\beta$)}
        \vspace{-3mm}
    \label{tab:estimate_drift}
    \centering
    \begin{tabular}[t]{llrrrrrr}
        \hline
        Country & Sex & Estimate & Lower C.I. & Upper C.I. & P-direction & $p$ & P-MAP\\
        \hline
Australia & Male & 0.0035 & -0.0100 & 0.0168 & 0.669 & 0.662 & 0.8556\\
 & Female & 0.0031 & -0.0081 & 0.0137 & 0.710 & 0.580 & 0.8282\\
Canada & Male & 0.0036 & -0.0060 & 0.0139 & 0.713 & 0.574 & 0.7857\\
& Female & 0.0035 & -0.0064 & 0.0137 & 0.785 & 0.430 & 0.7121\\
Denmark & Male & 0.0016 & -0.0053 & 0.0100 & 0.638 & 0.724 & 0.9306\\
 & Female & 0.0022 & -0.0046 & 0.0104 & 0.748 & 0.504 & 0.7661\\
England \& Wales & Male & 0.0030 & -0.0069 & 0.0149 & 0.688 & 0.624 & 0.9182\\
 & Female & 0.0030 & -0.0075 & 0.0139 & 0.716 & 0.568 & 0.8258\\
Finland & Male & 0.0026 & -0.0087 & 0.0154 & 0.663 & 0.674 & 0.9206\\
 & Female & 0.0037 & -0.0057 & 0.0152 & 0.777 & 0.446 & 0.7542\\
France & Male & 0.0139 & -0.0029 & 0.0307 & 0.936 & 0.128 & 0.3514\\
 & Female & 0.0102 & -0.0048 & 0.0241 & 0.905 & 0.190 & 0.3825\\
Italy & Male & 0.0091 & -0.0086 & 0.0251 & 0.819 & 0.362 & 0.7094\\
 & Female & 0.0034 & -0.0135 & 0.0264 & 0.697 & 0.606 & 0.9007\\
Japan & Male & 0.0035 & -0.0177 & 0.0231 & 0.612 & 0.776 & 0.9285\\
 & Female & -0.0009 & -0.0166 & 0.0181 & 0.508 & 0.984 & 0.9765\\
Netherlands & Male & 0.0040 & -0.0065 & 0.0142 & 0.778 & 0.444 & 0.7263\\
& Female & 0.0071 & -0.0030 & 0.0173 & 0.910 & 0.180 & 0.4226\\
Norway & Male & 0.0011 & -0.0062 & 0.0092 & 0.662 & 0.676 & 0.9616\\
& Female & 0.0024 & -0.0050 & 0.0095 & 0.740 & 0.520 & 0.7428\\
Sweden & Male & 0.0015 & -0.0052 & 0.0093 & 0.623 & 0.754 & 0.9727\\
 & Female & 0.0023 & -0.0052 & 0.0097 & 0.755 & 0.490 & 0.8701\\
United States & Male & 0.0042 & -0.0090 & 0.0146 & 0.753 & 0.494 & 0.7368\\
 & Female & 0.0018 & -0.0094 & 0.0143 & 0.660 & 0.680 & 0.8562\\

        \hline

        \multicolumn{8}{l}{The $p$ values are derived from P-direction.}
        
    \end{tabular}
    
\end{table}

\begin{table}[H]
    \caption{Minimum detectable drift ($\beta_{MDD}$)}
        \vspace{-3mm}
    \label{tab:MDD}
    \centering
    \begin{tabular}[t]{llrrr}
        \hline
        Country & Sex & MDD(\%) & Lower C.I. & Upper C.I.\\
        \hline
Australia & Male & 1.5291 & 1.1741 & 2.0575\\
 & Female & 1.3302 & 1.0383 & 1.7304\\
Canada & Male & 1.2660 & 0.9575 & 1.6411\\
 & Female & 1.2019 & 0.9210 & 1.6002\\
Denmark & Male & 0.8542 & 0.6710 & 1.1007\\
 & Female & 0.8310 & 0.6652 & 1.1033\\
England \& Wales & Male & 1.3872 & 1.1263 & 1.7274\\
 & Female & 1.3424 & 1.1015 & 1.6015\\
Finland & Male & 1.1916 & 0.9224 & 1.6462\\
 & Female & 1.1808 & 0.9122 & 1.5450\\
France & Male & 2.1546 & 1.7760 & 2.4512\\
 & Female & 1.9677 & 1.6538 & 2.3281\\
Italy & Male & 2.4857 & 2.0687 & 3.0534\\
 & Female & 2.6734 & 2.2186 & 3.1742\\
Japan & Male & 2.6183 & 2.0894 & 3.5640\\
& Female & 2.2731 & 1.7964 & 3.1221\\
Netherlands & Male & 1.2189 & 0.9540 & 1.4983\\
 & Female & 1.2019 & 0.9846 & 1.5282\\
Norway & Male & 0.9059 & 0.7116 & 1.1634\\
& Female & 0.7952 & 0.6384 & 1.0268\\
Sweden & Male & 0.7555 & 0.5924 & 0.9218\\
 & Female & 0.7880 & 0.6494 & 0.9518\\
United States & Male & 1.6251 & 1.2017 & 2.0115\\
 & Female & 1.6456 & 1.3319 & 2.0703\\
        \hline
    \end{tabular}
\end{table}

\begin{table}[H]
    \caption{Estimate and 95\% credible interval for magnitude of the period effects ($\sigma_{rw}$)}
            \vspace{-3mm}
    \label{tab:estimate_period_effect}
    \centering
    \begin{tabular}[t]{llrrr}
        \hline
        Country & Sex & Estimate & Lower C.I. & Upper C.I.\\
        \hline
            Australia & Male & 0.0501 & 0.0391 & 0.0681\\
& Female & 0.0447 & 0.0346 & 0.0574\\
            Canada & Male & 0.0439 & 0.0330 & 0.0563\\
& Female & 0.0422 & 0.0317 & 0.0549\\
            Denmark & Male & 0.0400 & 0.0315 & 0.0515\\
& Female & 0.0400 & 0.0312 & 0.0516\\
            England \& Wales & Male & 0.0646 & 0.0524 & 0.0801\\
& Female & 0.0622 & 0.0512 & 0.0743\\
            Finland & Male & 0.0488 & 0.0372 & 0.0661\\
& Female & 0.0489 & 0.0368 & 0.0621\\
            France & Male & 0.1064 & 0.0887 & 0.1220\\
& Female & 0.0975 & 0.0826 & 0.1159\\
            Italy & Male & 0.1037 & 0.0868 & 0.1275\\
& Female & 0.1114 & 0.0930 & 0.1324\\
            Japan & Male & 0.0765 & 0.0601 & 0.1019\\
& Female & 0.0679 & 0.0518 & 0.0894\\
            Netherlands & Male & 0.0537 & 0.0424 & 0.0664\\
& Female & 0.0536 & 0.0437 & 0.0677\\
            Norway & Male & 0.0413 & 0.0327 & 0.0533\\
& Female & 0.0372 & 0.0293 & 0.0471\\
            Sweden & Male & 0.0423 & 0.0342 & 0.0531\\
& Female & 0.0454 & 0.0374 & 0.0548\\
            United States & Male & 0.0526 & 0.0388 & 0.0647\\
 & Female & 0.0527 & 0.0430 & 0.0665\\
            
        \hline
    \end{tabular}
\end{table}

\begin{sidewaystable}
\begin{table}[H]
    \caption{Estimate for different starting age}
            \vspace{-3mm}
    \label{tab:estimate_different_starting_age}
    \centering
    \resizebox{\linewidth}{!}{%
    \begin{tabular}[t]{llcrrrrrrrrrrrrr}
        \hline
\toprule
& & & \multicolumn{3}{c}{Individual rate of aging ($b$)} 
  & \multicolumn{6}{c}{Drift ($\beta$)} 
  & \multicolumn{3}{c}{Magnitude of the period effects ($\sigma_{rw}$)} \\
\cmidrule(lr){4-6} \cmidrule(lr){7-12} \cmidrule(lr){13-15}
Country & Sex & Starting age
& Estimate & Lower C.I. & Upper C.I.
& Estimate & Lower C.I. & Upper C.I. & P-direction & $p$ & P-MAP
& Estimate & Lower C.I. & Upper C.I.  \\
        \hline
Denmark & Female & 50 & 0.0929 & 0.0919 & 0.0940 & 0.0012 & -0.0020 & 0.0040 & 0.578 & 0.711 & 0.8557 & 0.0182 & 0.0148 & 0.0219\\

&  & 60 & 0.0972 & 0.0956 & 0.0987 & 0.0003 & -0.0034 & 0.0043 & 0.928 & 0.536 & 0.9985 & 0.0223 & 0.0181 & 0.0267\\

&  & 70 & 0.1025 & 0.0983 & 0.1054 & 0.0010 & -0.0048 & 0.0055 & 0.758 & 0.621 & 0.9087 & 0.0285 & 0.0234 & 0.0357\\

 & Male & 50 & 0.0893 & 0.0868 & 0.0913 & 0.0006 & -0.0029 & 0.0044 & 0.776 & 0.612 & 0.9744 & 0.0206 & 0.0168 & 0.0246\\

&  & 60 & 0.0867 & 0.0836 & 0.0901 & 0.0001 & -0.0047 & 0.0046 & 0.916 & 0.542 & 0.9935 & 0.0238 & 0.0195 & 0.0289\\

&  & 70 & 0.0893 & 0.0865 & 0.0936 & 0.0001 & -0.0060 & 0.0053 & 0.888 & 0.556 & 0.9706 & 0.0308 & 0.0248 & 0.0388\\

England \& Wales & Female & 50 & 0.0784 & 0.0769 & 0.0789 & 0.0015 & -0.0015 & 0.0050 & 0.274 & 0.863 & 0.6800 & 0.0203 & 0.0174 & 0.0249\\

&  & 60 & 0.0937 & 0.0920 & 0.0946 & 0.0025 & -0.0025 & 0.0081 & 0.340 & 0.830 & 0.7057 & 0.0284 & 0.0248 & 0.0343\\

&  & 70 & 0.0985 & 0.0965 & 0.0996 & 0.0007 & -0.0061 & 0.0091 & 0.762 & 0.619 & 0.9472 & 0.0410 & 0.0340 & 0.0488\\

 & Male & 50 & 0.0760 & 0.0750 & 0.0767 & -0.0003 & -0.0037 & 0.0027 & 0.794 & 0.603 & 0.9632 & 0.0198 & 0.0171 & 0.0234\\

&  & 60 & 0.0874 & 0.0865 & 0.0882 & -0.0006 & -0.0061 & 0.0049 & 0.758 & 0.621 & 0.9528 & 0.0306 & 0.0267 & 0.0374\\

&  & 70 & 0.0873 & 0.0813 & 0.0908 & -0.0012 & -0.0084 & 0.0074 & 0.788 & 0.606 & 0.9347 & 0.0456 & 0.0370 & 0.0535\\

France & Female & 50 & 0.0846 & 0.0842 & 0.0849 & 0.0023 & -0.0002 & 0.0043 & 0.068 & 0.966 & 0.2310 & 0.0166 & 0.0141 & 0.0190\\

&  & 60 & 0.0934 & 0.0918 & 0.0954 & 0.0018 & -0.0023 & 0.0060 & 0.406 & 0.797 & 0.6775 & 0.0270 & 0.0237 & 0.0319\\

&  & 70 & 0.1029 & 0.0996 & 0.1061 & 0.0030 & -0.0040 & 0.0115 & 0.368 & 0.816 & 0.6648 & 0.0479 & 0.0398 & 0.0550\\

 & Male & 50 & 0.0806 & 0.0789 & 0.0810 & 0.0005 & -0.0018 & 0.0026 & 0.788 & 0.606 & 0.9885 & 0.0155 & 0.0135 & 0.0185\\

&  & 60 & 0.0874 & 0.0864 & 0.0889 & 0.0011 & -0.0028 & 0.0047 & 0.636 & 0.682 & 0.8312 & 0.0259 & 0.0224 & 0.0308\\

&  & 70 & 0.0939 & 0.0926 & 0.0952 & 0.0021 & -0.0071 & 0.0090 & 0.660 & 0.670 & 0.9055 & 0.0497 & 0.0430 & 0.0585\\

Italy & Female & 50 & 0.0922 & 0.0912 & 0.0933 & 0.0029 & -0.0034 & 0.0092 & 0.358 & 0.821 & 0.6206 & 0.0399 & 0.0331 & 0.0476\\

&  & 60 & 0.1032 & 0.1016 & 0.1049 & 0.0040 & -0.0037 & 0.0140 & 0.346 & 0.827 & 0.6820 & 0.0719 & 0.0594 & 0.0858\\

&  & 70 & 0.1036 & 0.0993 & 0.1053 & 0.0088 & -0.0068 & 0.0213 & 0.336 & 0.832 & 0.5521 & 0.1128 & 0.0945 & 0.1377\\

 & Male & 50 & 0.0859 & 0.0855 & 0.0863 & -0.0001 & -0.0045 & 0.0047 & 0.976 & 0.512 & 0.9904 & 0.0289 & 0.0236 & 0.0350\\

&  & 60 & 0.0935 & 0.0928 & 0.0943 & 0.0014 & -0.0061 & 0.0083 & 0.754 & 0.623 & 0.9346 & 0.0493 & 0.0415 & 0.0609\\

&  & 70 & 0.0961 & 0.0928 & 0.0992 & 0.0055 & -0.0092 & 0.0184 & 0.454 & 0.773 & 0.6689 & 0.0860 & 0.0703 & 0.1039\\

Norway & Female & 50 & 0.0864 & 0.0854 & 0.0877 & 0.0020 & -0.0020 & 0.0062 & 0.358 & 0.821 & 0.6747 & 0.0238 & 0.0193 & 0.0294\\

&  & 60 & 0.0933 & 0.0917 & 0.1030 & 0.0015 & -0.0034 & 0.0058 & 0.494 & 0.753 & 0.7616 & 0.0252 & 0.0204 & 0.0316\\

&  & 70 & 0.1012 & 0.0989 & 0.1049 & 0.0030 & -0.0034 & 0.0094 & 0.354 & 0.823 & 0.6414 & 0.0342 & 0.0267 & 0.0410\\

 & Male & 50 & 0.0845 & 0.0830 & 0.0858 & 0.0008 & -0.0038 & 0.0051 & 0.670 & 0.665 & 0.9663 & 0.0239 & 0.0194 & 0.0289\\

&  & 60 & 0.0910 & 0.0897 & 0.0921 & 0.0002 & -0.0040 & 0.0054 & 0.832 & 0.584 & 0.9996 & 0.0240 & 0.0195 & 0.0293\\

&  & 70 & 0.0974 & 0.0928 & 0.1012 & 0.0003 & -0.0065 & 0.0071 & 0.914 & 0.543 & 0.9979 & 0.0332 & 0.0261 & 0.0414\\

Sweden & Female & 50 & 0.0902 & 0.0887 & 0.0913 & 0.0014 & -0.0007 & 0.0042 & 0.234 & 0.883 & 0.5209 & 0.0171 & 0.0148 & 0.0201\\

&  & 60 & 0.1026 & 0.1004 & 0.1040 & 0.0005 & -0.0028 & 0.0047 & 0.728 & 0.636 & 0.9328 & 0.0284 & 0.0244 & 0.0330\\

&  & 70 & 0.1053 & 0.1026 & 0.1087 & 0.0022 & -0.0036 & 0.0075 & 0.440 & 0.780 & 0.6943 & 0.0381 & 0.0316 & 0.0445\\

 & Male & 50 & 0.0727 & 0.0719 & 0.0755 & 0.0011 & -0.0016 & 0.0037 & 0.392 & 0.804 & 0.6968 & 0.0167 & 0.0143 & 0.0194\\

&  & 60 & 0.0946 & 0.0935 & 0.0957 & 0.0006 & -0.0030 & 0.0041 & 0.806 & 0.597 & 0.9911 & 0.0238 & 0.0202 & 0.0282\\

&  & 70 & 0.0974 & 0.0943 & 0.1023 & 0.0004 & -0.0053 & 0.0056 & 0.916 & 0.542 & 0.9913 & 0.0372 & 0.0314 & 0.0452\\
\hline
    \end{tabular}
    }
\end{table}
\end{sidewaystable}

\clearpage   

\section*{Supplementary Results and Model Diagnostics}
\label{appendix:decomposition}

\begin{figure}[p]
\centering
\includegraphics[width=\textwidth]{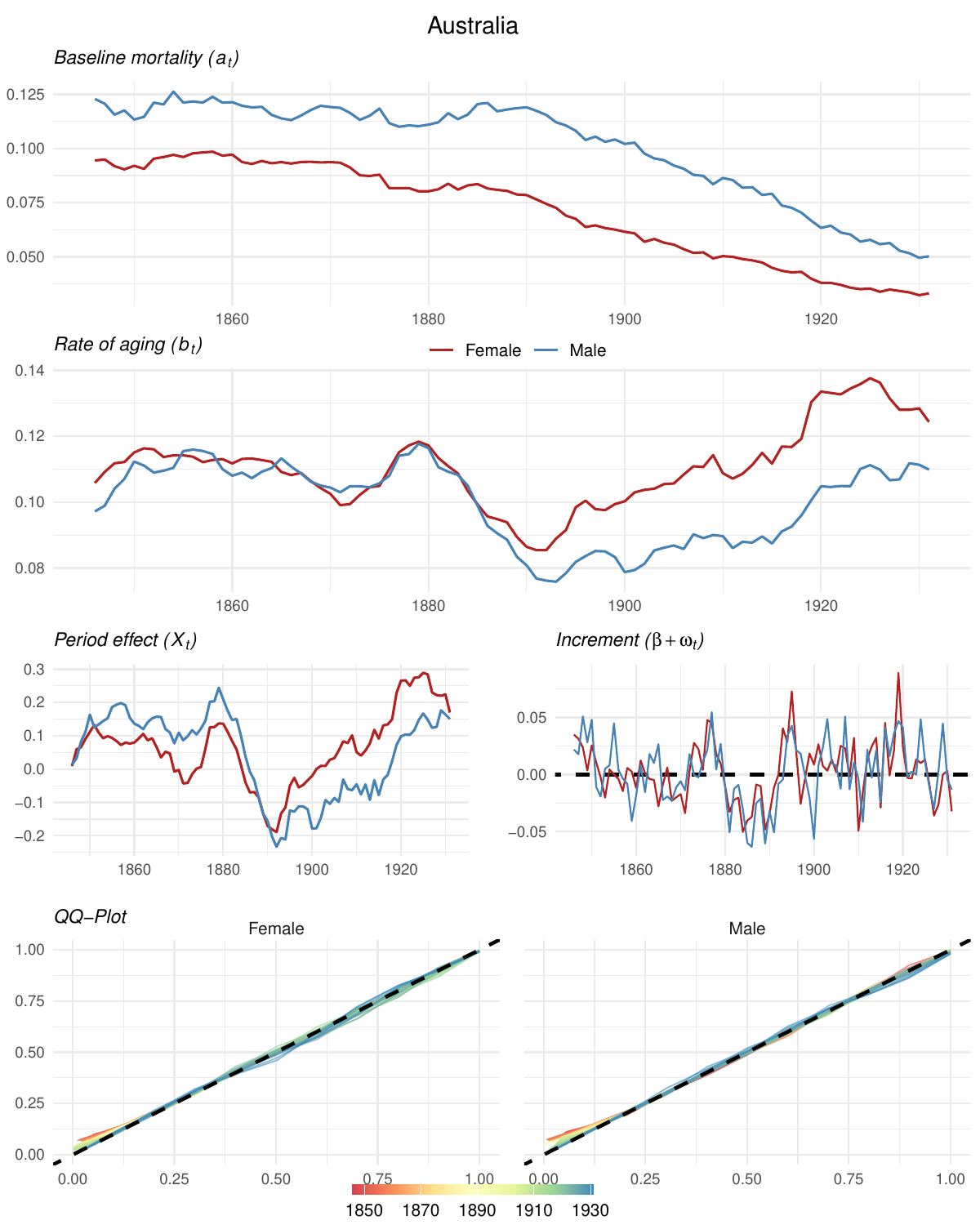} 
\caption{Random walk decomposition for Australia.}
\label{fig:Australia}
\end{figure}

\begin{figure}[p]
\centering
\includegraphics[width=\textwidth]{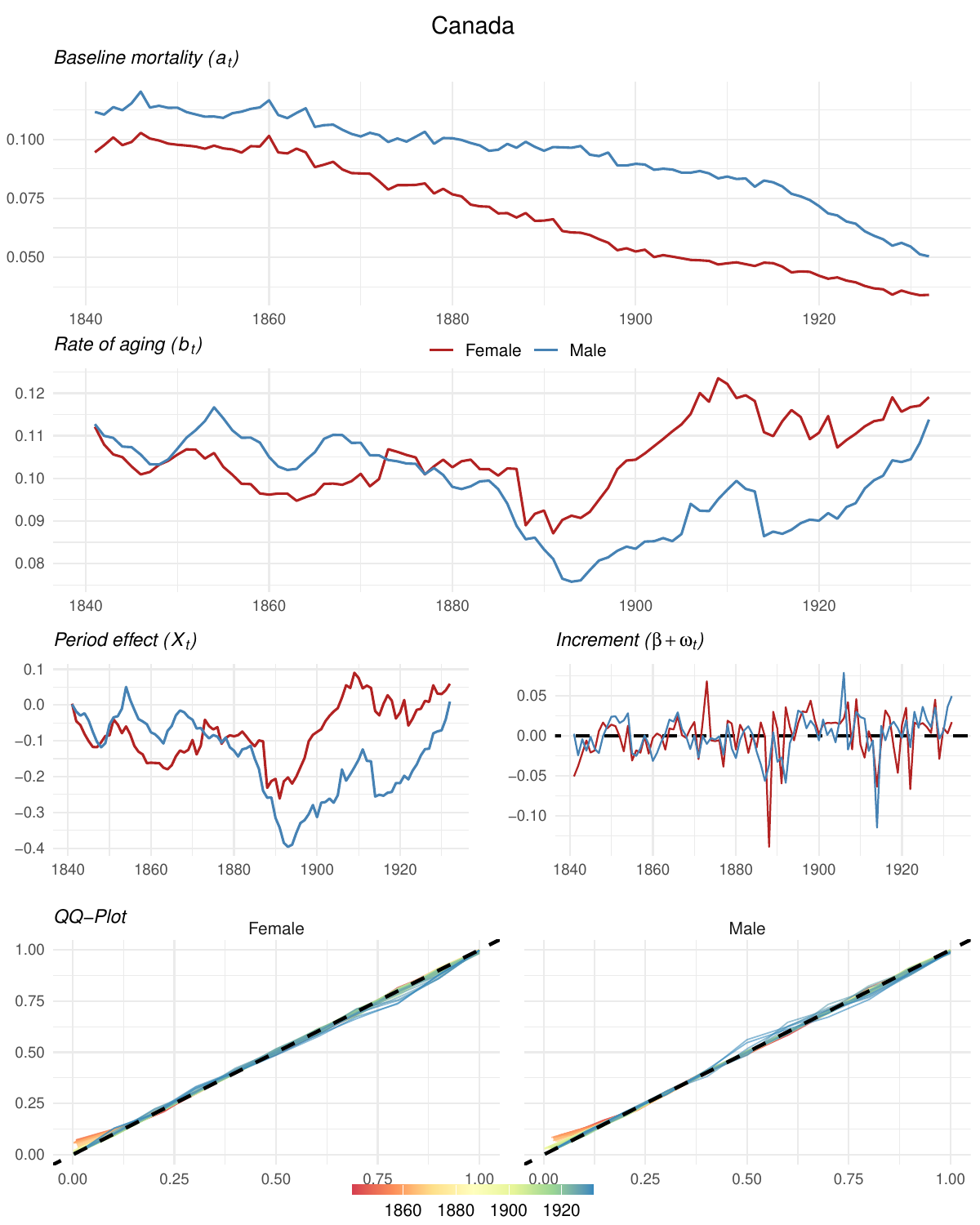} 
\caption{Random walk decomposition for Canada.}
\label{fig:Canada}
\end{figure}

\begin{figure}[p]
\centering
\includegraphics[width=\textwidth]{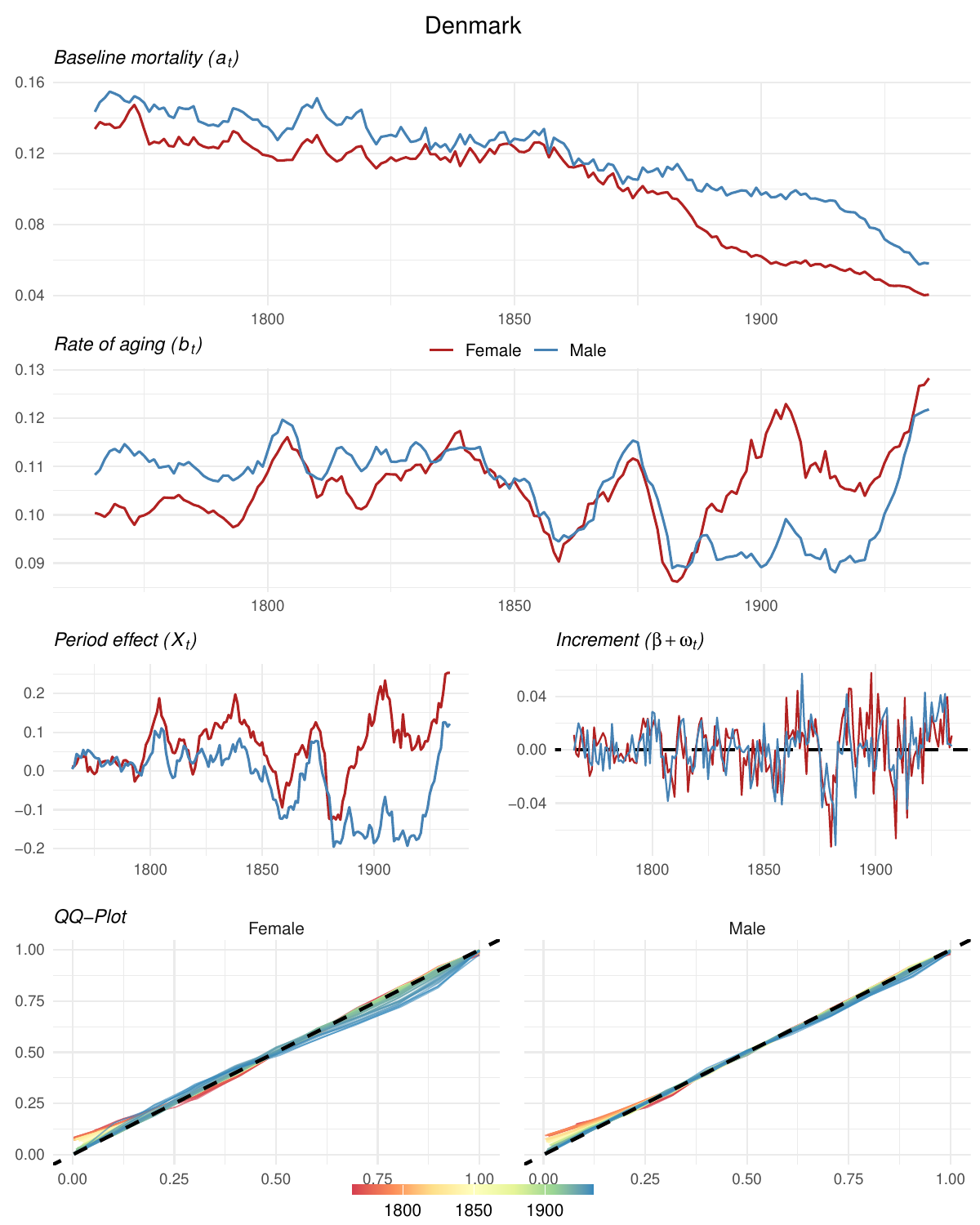} 
\caption{Random walk decomposition for Denmark.}
\label{fig:Denmark}
\end{figure}

\begin{figure}[p]
\centering
\includegraphics[width=\textwidth]{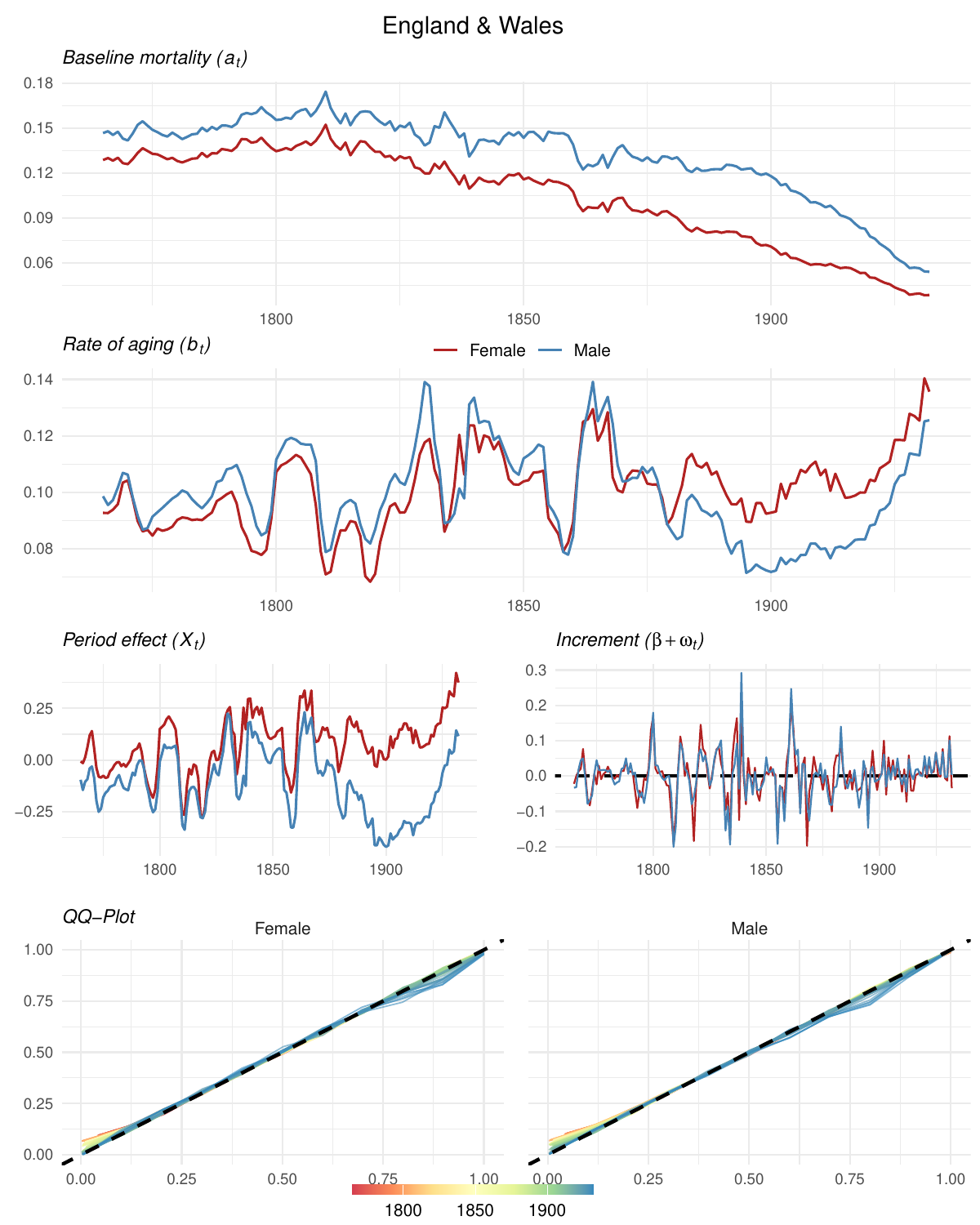} 
\caption{Random walk decomposition for England \& Wales.}
\label{fig:England}
\end{figure}

\begin{figure}[p]
\centering
\includegraphics[width=\textwidth]{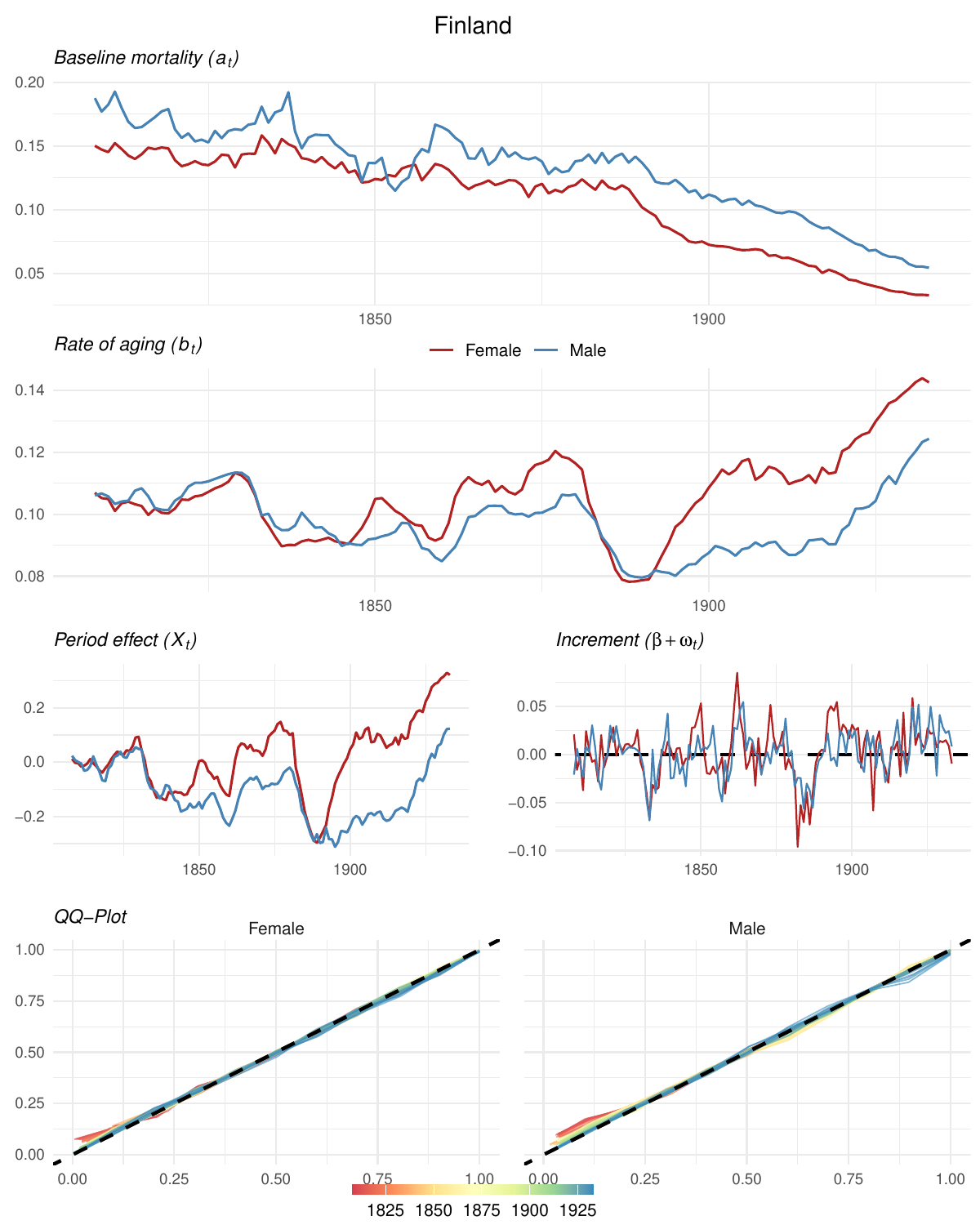} 
\caption{Random walk decomposition for Finland.}
\label{fig:Finland}
\end{figure}

\begin{figure}[p]
\centering
\includegraphics[width=\textwidth]{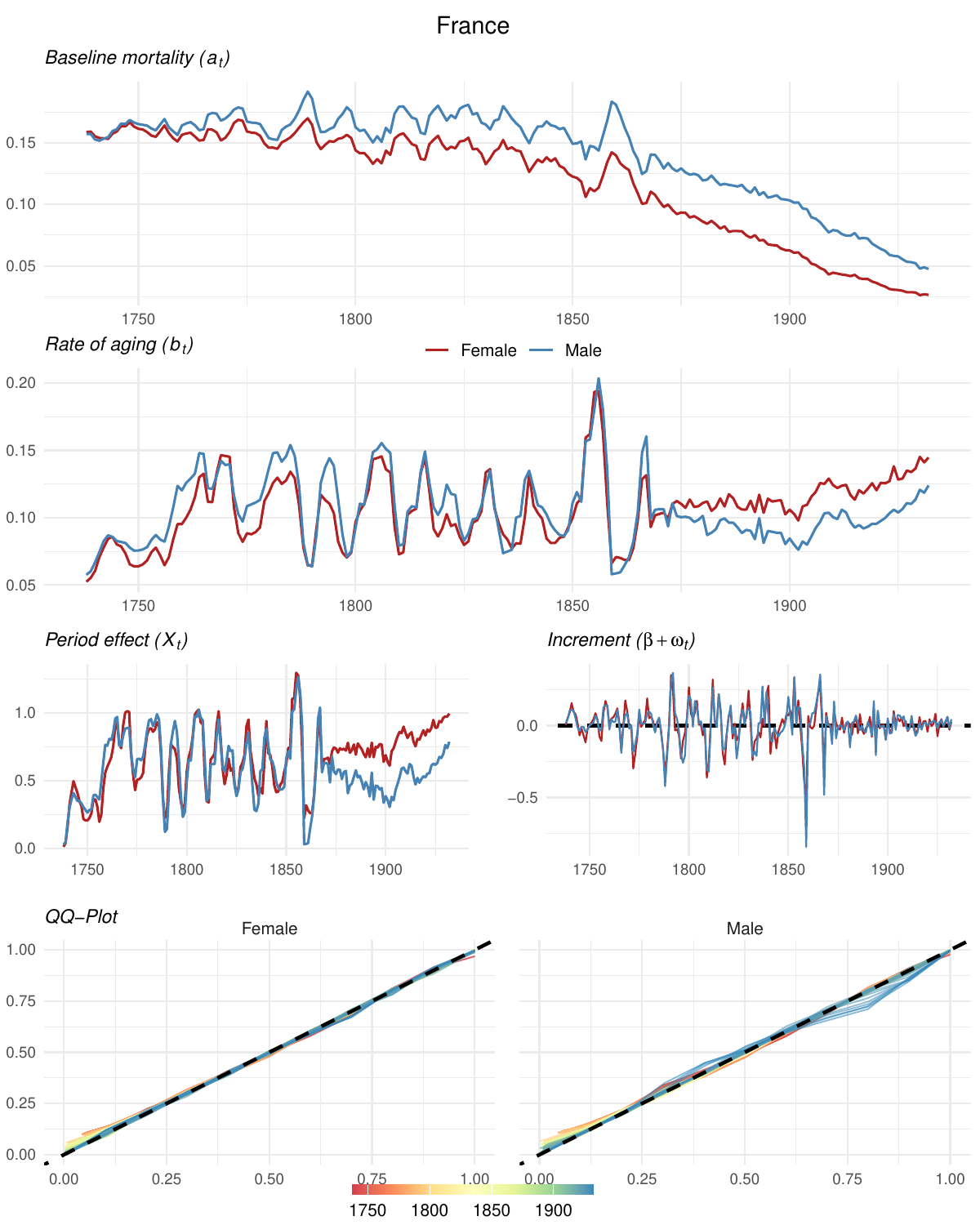} 
\caption{Random walk decomposition for France.}
\label{fig:France}
\end{figure}

\begin{figure}[p]
\centering
\includegraphics[width=\textwidth]{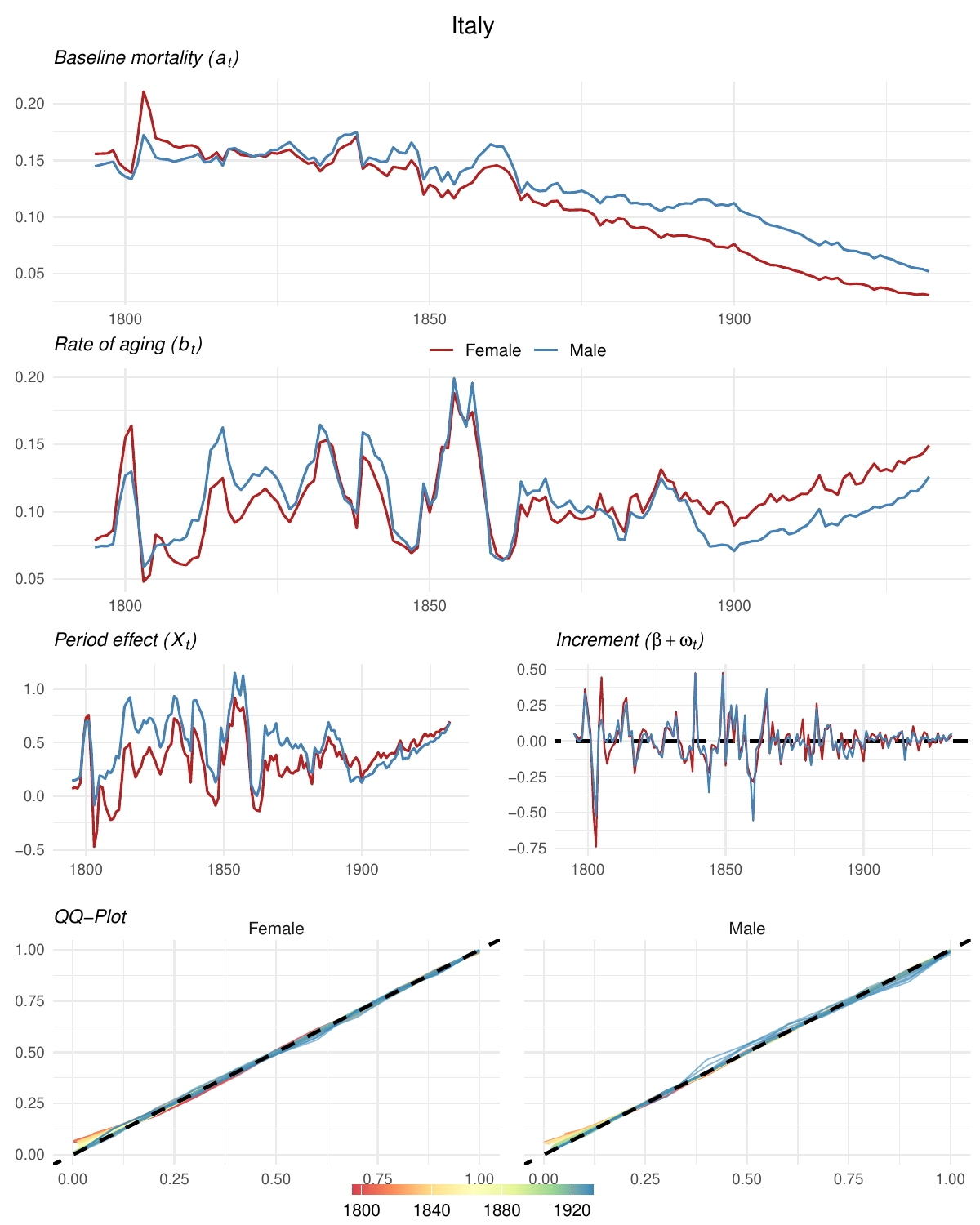} 
\caption{Random walk decomposition for Italy.}
\label{fig:Italy}
\end{figure}

\begin{figure}[p]
\centering
\includegraphics[width=\textwidth]{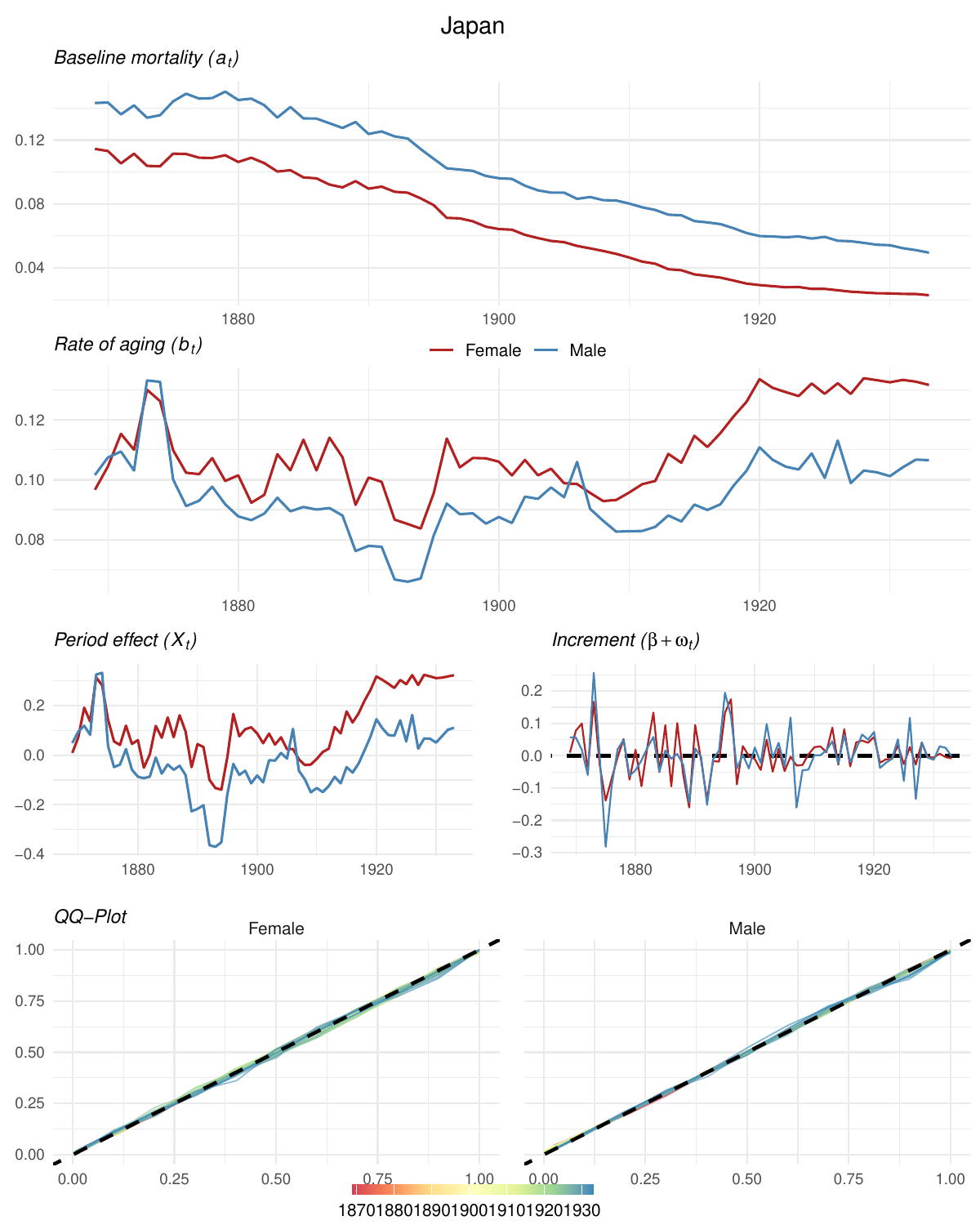} 
\caption{Random walk decomposition for Japan.}
\label{fig:Japan}
\end{figure}

\begin{figure}[p]
\centering
\includegraphics[width=\textwidth]{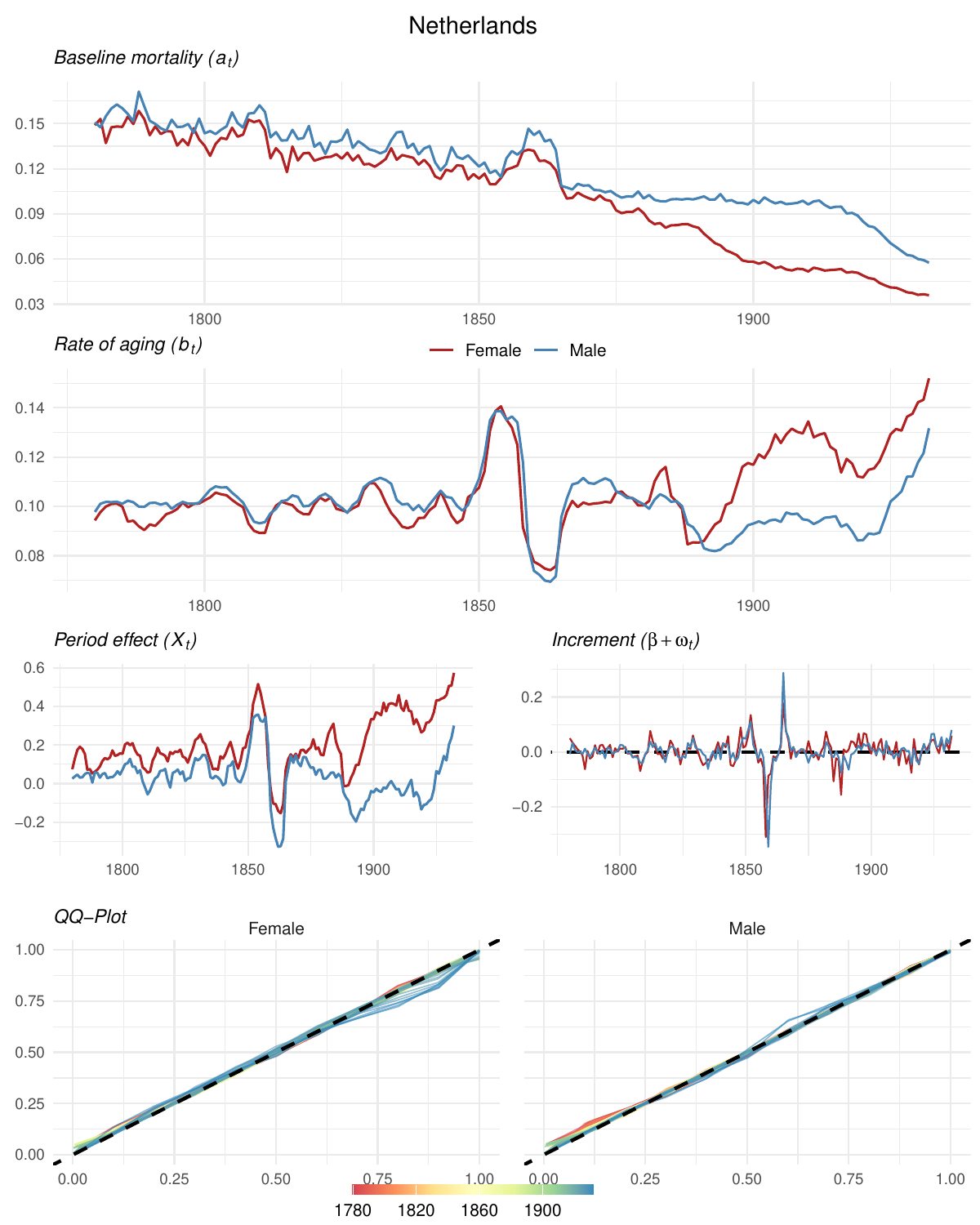} 
\caption{Random walk decomposition for Netherlands.}
\label{fig:Netherlands}
\end{figure}

\begin{figure}[p]
\centering
\includegraphics[width=\textwidth]{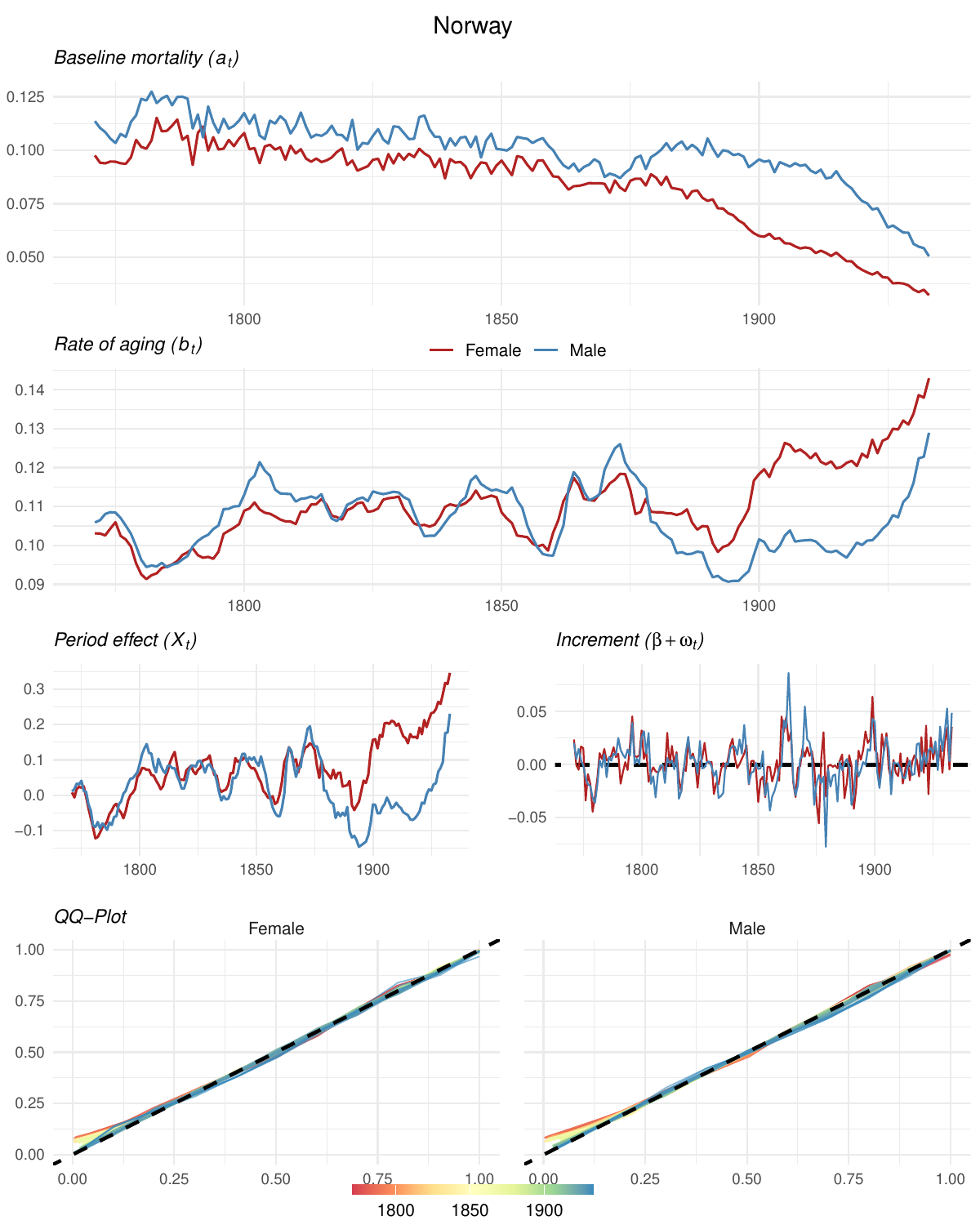} 
\caption{Random walk decomposition for Norway.}
\label{fig:Norway}
\end{figure}

\begin{figure}[p]
\centering
\includegraphics[width=\textwidth]{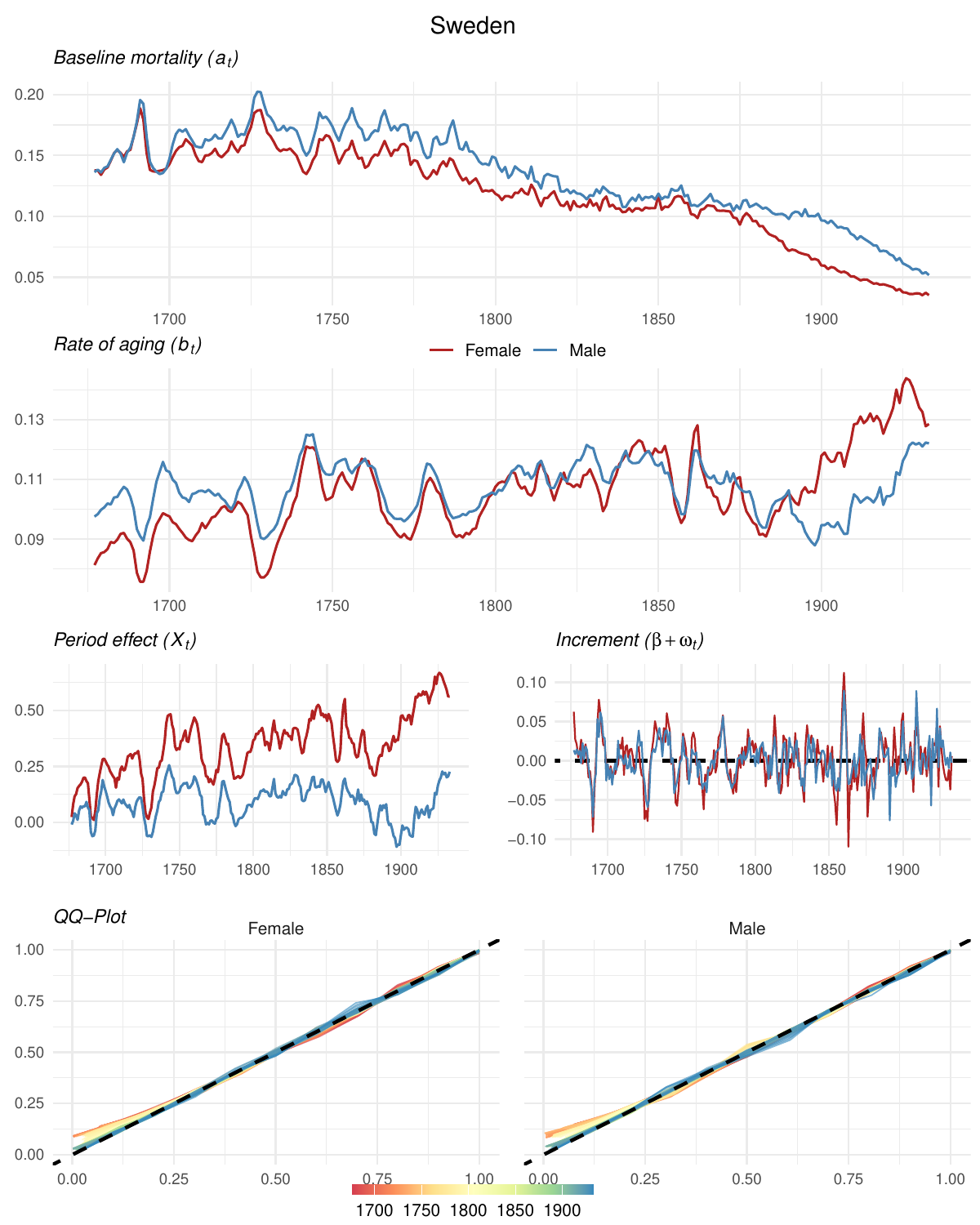} 
\caption{Random walk decomposition for Sweden.}
\label{fig:Sweden}
\end{figure}

\begin{figure}[p]
\centering
\includegraphics[width=\textwidth]{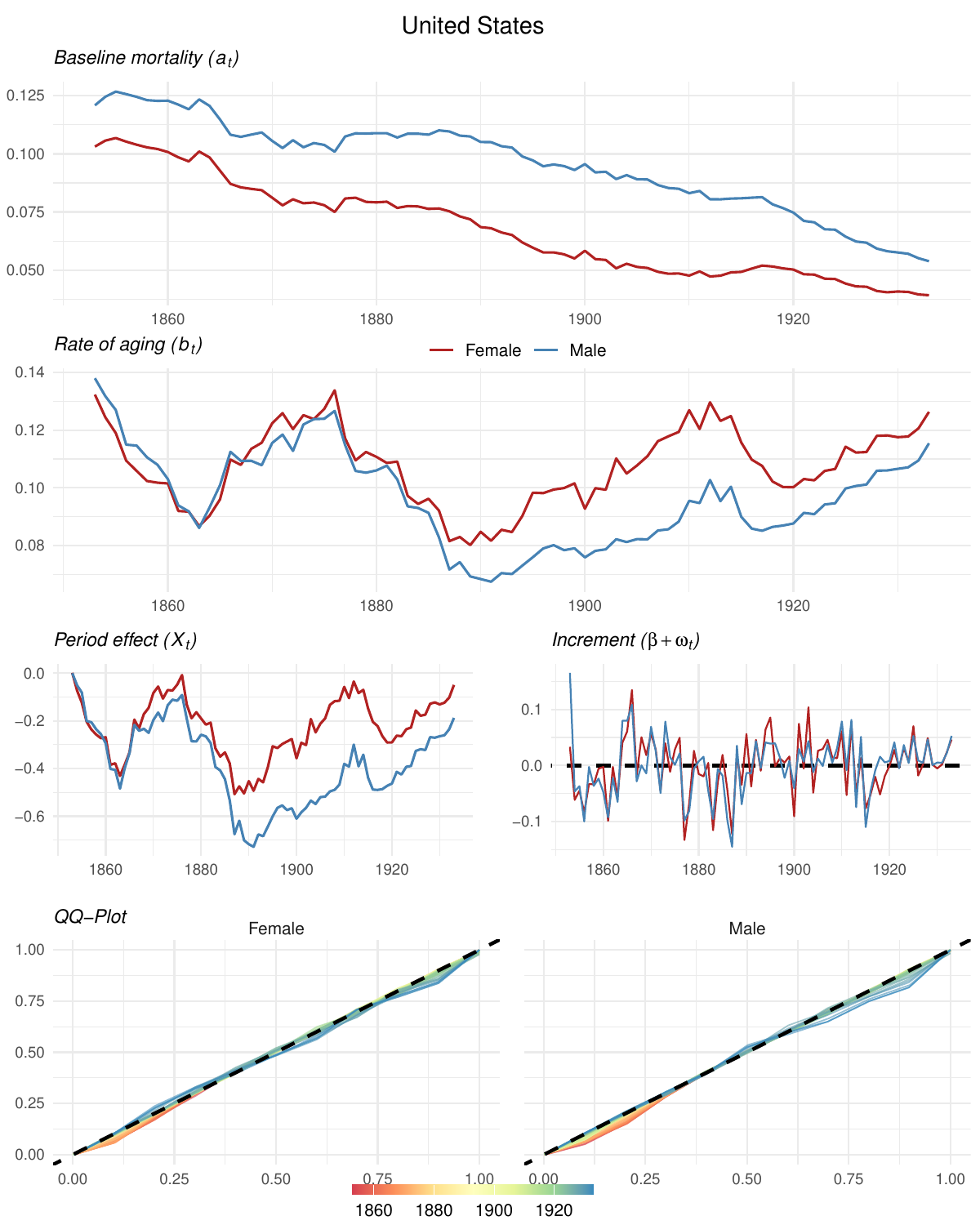} 
\caption{Random walk decomposition for United States.}
\label{fig:UnitedStates}
\end{figure}

\begin{figure}[p]
\centering
\includegraphics[width=\textwidth]{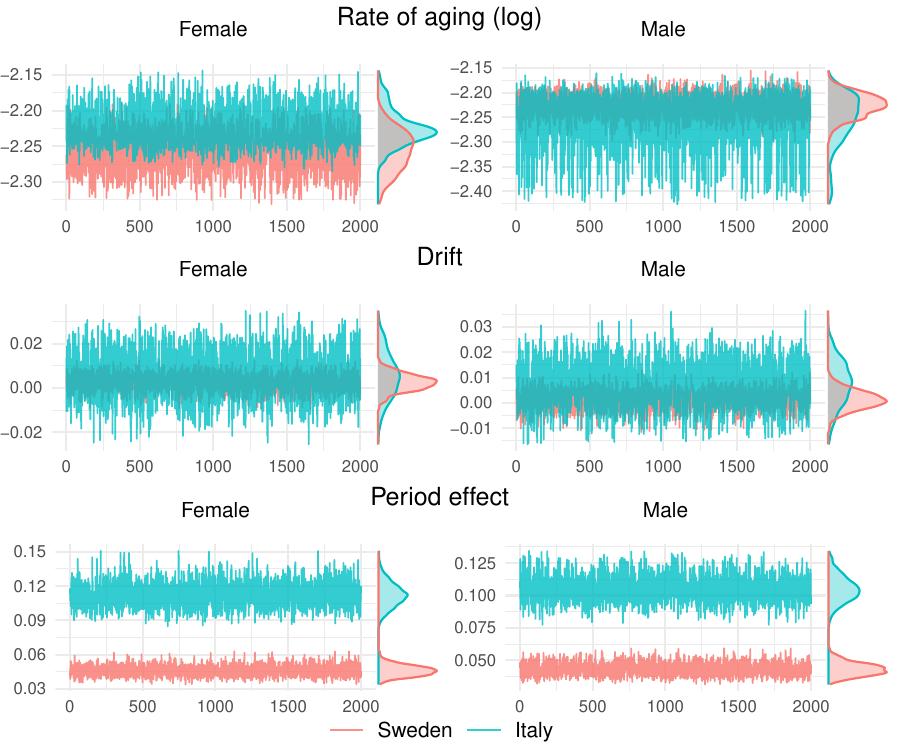} 
\caption{Trace plots for the baseline rate ($\log b$), drift ($\beta$), and period effect ($\sigma_{\text{rw}}$) for Italy and Sweden.}
\label{fig:traceplots}
\end{figure}

This section provides the detailed country-by-country results of the random walk decomposition model. For each of the 12 countries, we present a five-panel figure presenting: the estimated series of the baseline mortality ($a_t$) and the rate of aging ($b_t$); the estimated latent period effect, $X_t$; the estimated cohort increments, $\Delta X_t = b_t - b_{t-1}$; and a posterior predictive QQ-plot for model validation (Figure~\ref{fig:Australia}-~\ref{fig:UnitedStates}). To supplement the $\hat{R}$ statistics reported in the main text, Figure~\ref{fig:traceplots} displays the MCMC trace plots for the main parameters for Italy (a high-volatility country) and Sweden (a low-volatility country). The chains are well-mixed and stationary, providing a visual check of convergency.



\clearpage

\end{document}